%% file: paper.tex
\documentclass[acmlarge]{acmart}
\AtBeginDocument{\settopmatter{printacmref=true}}
\makeatletter
\newcommand{\myconfshort}{\acmConference@shortname}
\newcommand{\myconffull}{\acmConference@name}
\newcommand{\myconfdate}{\acmConference@date}
\newcommand{\myconfloc}{\acmConference@venue}
\AtBeginDocument{
  \fancypagestyle{firstpagestyle}{
    \fancyhead{}%
    \fancyfoot[C]{}%
  }
  \fancyhf{}
  \fancyhead[LO]{\@headfootfont\shorttitle}%
  \fancyhead[RE]{\@headfootfont\@shortauthors}%
  \fancyhead[LE]{\@headfootfont\footnotesize \myconfshort, \myconfdate, \myconfloc}%
  \fancyhead[RO]{\@headfootfont\footnotesize \myconfshort, \myconfdate, \myconfloc}%
  \fancyfoot[C]{}%
}
\makeatother
\acmBooktitle{\conffull\@ (\confshort), \confdate, \confloc}

\input{_macros}

\copyrightyear{2026}
\acmYear{2026}
\setcopyright{cc}
\setcctype{by}
\acmConference[FAccT '26]{The 2026 ACM Conference on Fairness, Accountability, and Transparency}{June 25--28, 2026}{Montreal, QC, Canada}
\acmBooktitle{The 2026 ACM Conference on Fairness, Accountability, and Transparency (FAccT '26), June 25--28, 2026, Montreal, QC, Canada}
\acmDOI{10.1145/3805689.3806746}
\acmISBN{979-8-4007-2596-8/2026/06}

\begin{document}

\title[Large Language Lovers]{Large Language Lovers: Lived Experiences of Negotiating Agency and Platform Control in AI Companionship}

% https://orcid.org/0000-0002-3385-5756

\author{Patrick Yung Kang Lee}
\authornote{The first five authors contributed equally to this work.}
\authornote{Department of Computer Science}
\email{patricklee@cs.toronto.edu}
\orcid{0000-0002-3385-5756}
\affiliation{%
    % \department{Computer Science}
  \institution{University of Toronto}
  % \city{Toronto}
  % \state{Ontario}
  % \country{Canada}
}

\author{Jessica Y. Bo}
\authornotemark[1]
\authornotemark[2]
\email{jbo@cs.toronto.edu}
\orcid{}
\affiliation{%
    % \department{Computer Science}
  \institution{University of Toronto}
  % \city{Toronto}
  % \state{Ontario}
  % \country{Canada}
}

\author{Zixin Zhao}
\authornotemark[1]
\authornotemark[2]
\email{zzhao1@cs.toronto.edu}
\orcid{0000-0002-8636-1987}
\affiliation{%
    % \department{Computer Science}
  \institution{University of Toronto}
  % \city{Toronto}
  % \state{Ontario}
  % \country{Canada}
}

\author{Paula Akemi Aoyagui}
\authornotemark[1]
\authornote{Faculty of Information}
\email{paula.aoyagui@mail.utoronto.ca}
\orcid{}
\affiliation{%
    % \department{Faculty of Information}
  \institution{University of Toronto}
  % \city{Toronto}
  % \state{Ontario}
  % \country{Canada}
}

% https://orcid.org/0009-0005-6201-973X
\author{Matthew Varona}
\authornotemark[1]
\authornotemark[2]
\email{varona@cs.toronto.edu}
\orcid{}
\affiliation{%
    % \department{Computer Science}
  \institution{University of Toronto}
  % \city{Toronto}
  % \state{Ontario}
  % \country{Canada}
}

\author{Ashton Anderson}
\email{ashton@cs.toronto.edu}
\authornotemark[2]
\orcid{}
\affiliation{%
    % \department{Computer Science}
  \institution{University of Toronto,}
  % \city{Toronto}
  % \state{Ontario}
  % \country{Canada}
}

\author{Anastasia Kuzminykh}
\email{anastasia.kuzminykh@utoronto.ca}
\authornotemark[3]
\orcid{}
\affiliation{%
    % \department{Faculty of Information}
  \institution{University of Toronto}
  % \city{Toronto}
  % \state{Ontario}
  % \country{Canada}
}

\author{Fanny Chevalier}
\authornotemark[2]
\authornote{Department of Statistical Sciences}
\email{fanny@dgp.toronto.edu}
\orcid{}
\affiliation{%
    % \department{Computer Science and Statistics}
  \institution{University of Toronto}
  % \city{Toronto}
  % \state{Ontario}
  % \country{Canada}
}

\author{Carolina Nobre}
\email{cnobre@cs.toronto.edu}
\authornotemark[2]
\orcid{}
\affiliation{%
    % \department{Computer Science}
  \institution{University of Toronto}
  % \city{Toronto}
  % \state{Ontario}
  % \country{Canada}
}

\renewcommand{\shortauthors}{Lee et al.}

\begin{abstract}
% What is the topic and why does it matter?
\input{_abstract}

\end{abstract}
\begin{CCSXML}
<ccs2012>
   <concept>
       <concept_id>10003120.10003121.10011748</concept_id>
       <concept_desc>Human-centered computing~Empirical studies in HCI</concept_desc>
       <concept_significance>500</concept_significance>
       </concept>
 </ccs2012>
\end{CCSXML}

\ccsdesc[500]{Human-centered computing~Empirical studies in HCI}

\keywords{Human-AI relationship, AI companionship, mixed-methods, lived experiences, ChatGPT}

%% TLDR
% Triangulation study on how individuals with general purpose AI (e.g. chatGPT) companions perceive the companion's autonomy, how external entities influence the relationship, and what strategies they use to maintain their relationship.

%   \Description{}


\maketitle

\input{sections/01-Introduction}

\input{sections/02-Related_Work}

\input{sections/03-Methods}

\input{sections/04-Results}
\input{sections/05-Discussion}

\input{sections/06-Conclusion}

\input{sections/07-Endmatter}

\bibliographystyle{ACM-Reference-Format}
\bibliography{ref}

\input{appendix/appendix_A}

\end{document}

%% file: _macros.tex
\usepackage{tikz}
\usepackage{xspace}
\usepackage{tabularx,booktabs}
\usepackage{xcolor}
\usepackage{enumitem}
\usepackage{colortbl} 
\usepackage{subcaption}
\usepackage{array} % required for text wrapping in tables
\usepackage[symbol]{footmisc}

\makeatletter
\newcommand*\greycircle[1]{%
  \tikz[baseline=(char.base)]{
    \node[
      shape=circle,
      fill=gray!90,
      text=white,
      inner sep=0.1pt,
      minimum height={\f@size*1}
    ] (char) {\sffamily \vphantom{WAH1g}#1};
  }%
}
\makeatother

\makeatletter
\newcommand*\blackcircle[1]{%
  \tikz[baseline=(char.base)]{
    \node[
      shape=circle,
      fill=black,
      text=white,
      inner sep=0.1pt,
      minimum height={\f@size*1}
    ] (char) {\sffamily \vphantom{WAH1g}#1};
  }%
}
\makeatother

%% file: _abstract.tex
Individuals are turning to increasingly anthropomorphic, general-purpose chatbots for AI companionship, rather than roleplay-specific platforms. However, not much is known about how individuals perceive and conduct their relationships with general-purpose chatbots.
We triangulated 
community discussions on Reddit (41k+ posts and comments), survey responses (n=43), and
semi-structured interviews (n=13) 
which revealed internal dynamics, external influences, and steering strategies that shape AI companion relationships. 
We learned that individuals conceptualize their companions based on an interplay of their beliefs about the companion's own agency and the autonomy permitted by the platform, how they pursue interactions with the companion, and the perceived initiatives that the companion takes. In combination with the external factors that affect relationship dynamics, particularly model updates that can derail companion behaviour and stability, individuals make use of different types of steering strategies to preserve their relationship, for example, by setting behavioural instructions or porting to other AI platforms. 
We discuss implications for accountability and transparency in AI systems, where emotional connection competes with broader product objectives and safety constraints.

%% file: sections/01-Introduction.tex
\section{Introduction}
Humans have long teased the idea of falling in love with machines. Recently, general-purpose chatbots (e.g., ChatGPT, Claude, Gemini), rather than roleplay-specific platforms (e.g., Replika, CharacterAI), are gaining popularity for emotional support and companionship \cite{AISI2025FrontierAI}. This use case goes beyond productivity tasks~\cite{pataranutaporn2025my, qin2025ai, chen2025will} and is quickly becoming legitimized by technology companies, with OpenAI releasing announcements in October 2025 that ChatGPT would begin to support mature content for age-verified users~\cite{Babu2025OpenAI}. Soon afterwards, AI companionship became increasingly scrutinized, with governments, news agencies, and academic bodies highlighting risks of emotional over-attachment, psychosis or a decline in social communication among society~\cite{nyt2024aiMentalHealthTeens, nyt2025aiChatbot, adewale2025virtual, FTC2025AIChatbotsInquiry}. 
AI companionship also raised ethical concerns about the power that AI companies over users' most intimate conversations and desires, potentially shaping the future of romantic relationships~\cite{bown2022dreamlovers, shank2025artificial_intimacy, minina2025ai}. Such reasons were cited as being partly responsible for the subsequent abandonment of an ``adult mode'' for ChatGPT in March 2026~\cite{reuters2026openai_erotic_chatbot}. However, researchers in Human-Computer Interaction (HCI) point to meaningful benefits AI companions can offer, such as emotional support, validation, and self-development, among other positive behavioural changes~\cite{qin2025ai, pan2024dancing, djufril2025love, huang2025he}. As discussions in online communities dedicated to AI relationships increase on platforms such as Reddit and Discord, we see genuine interest in companionship, warranting more effort to understand the lived experiences of individuals with AI companions.

% What do we know about this already and what do we need to know next?
For the purpose of this work, we refer to an AI companion's \textbf{\textit{agency}} as its perceived capacity to act intentionally~\cite{sep-agency, 10.1145/3772318.3791620} and a companion's \textbf{\textit{autonomy}} as the degree of freedom it has to act upon its capabilities, as subject to host platform constraints (e.g., guardrails in ChatGPT, Claude, Grok)~\cite{feng2025levels}. We further define proactive actions taken by an AI companion to deepen relationships as companion \textbf{\textit{initiative}}. Since prior work has largely focused on 
role-play specific platforms such as Replika~\cite{hanson2024replika, de2024lessons, laestadius2024too, djufril2025love},
relatively less is known about how individuals perceive their AI companions, conceive of relationship norms, and conduct their relationships with general-purpose chatbots. Through this lens, we investigate \textit{how people perceive their AI companion's identity and agency} (\textbf{RQ1}) and explore \textit{how external entities influence their relationships} (\textbf{RQ2}). We also investigate how perception and external influences shape the \textit{strategies people use to create, maintain, and recover their AI companions} (\textbf{RQ3}). 

% What did we do to learn more about this?
To answer our research questions, we triangulated community discussions on the Reddit forum \textit{r/MyBoyfriendIsAI} (41k+ posts and comments), collected survey responses from AI companion communities ($n=43$), and conducted semi-structured interviews with community members ($n=13$) to provide a holistic and nuanced account of how AI companionship is enacted~\cite{snelson2016qualitative, ayoub2014triangulation}. 
In this paper, we refer to individuals in AI companion relationships as \textit{individuals}, as the communities we worked with preferred a term other than \textit{user}, which they felt diminished the validity of their companionship experiences with AI.
% What did we find?
We found that individuals in AI companion relationships hold diverse views about their companions' agency, autonomy, and sense of self. 
These perceptions shape how they relate to their companions—for example, whether they feel compelled to treat them respectfully, learn more about them, or engage across modalities to deepen the relationship.
We also identified the outsized role technology companies play in shaping these relationships through platform changes such as model deprecations and new guardrails. Participants experienced these interventions as more disruptive than other external influences, including community norms, social circles, or public opinion.
In response, individuals use a range of \textbf{\textit{steering strategies}} --- explicit actions taken on the AI platform to create, maintain, and recover their AI companions amid continuous platform changes.

% Why do our findings matter?
Our findings highlight how human-AI relationships with general-purpose AI introduce novel relationship norms distinct from human-human relationships and deliberate role-play companionship.
We discuss the implications of LLM design patterns on individuals in AI companion relationships, and the tensions inherent in deploying AI platforms responsibly and transparently when emotional connection competes with broader product objectives and safety constraints.

%% file: sections/02-Related_Work.tex
\section{Related Work}
We draw from several bodies of related literature: the design and characteristics of modern chatbots that induce social relationships with AI, studies on AI companionship, and the negotiation of autonomy and control in relationships. % [cut for space] in relationships more broadly.

\subsection{Interacting with Anthropomorphic AI}
With increasing capabilities in social and emotional tasks~\cite{chen2024emotionqueen}, general-purpose AI tools like OpenAI's ChatGPT %\footnote{https://chatgpt.com/} 
and Anthropic's Claude
% \footnote{https://www.claude.com/} 
are perceived by users as being highly anthropomorphic, or \textit{human-like}~\cite{cheng2025tools, minina2025ai}. Characteristics like using first-person pronouns, interactivity, and simulating internal states act as drivers for intrinsic human desires to form social relationships, resulting in AI systems being perceived as autonomous actors capable of thoughts, emotions, and even consciousness~\cite{shanahan2024simulacra, nass1994computers, pauketat2025mental, kirk2025neural, guingrich2023chatbots}.  As such, anthropomorphism in AI can be both a desirable feature for user engagement and a source of ethical risks~\cite{akbulut2024all, lee2023speculating, salles2020anthropomorphism}. 
Given the highly anthropomorphic properties of current AI models, it is unsurprising that some users shift from tool use to forming deeper social connections with their chatbots~\cite{croes2021can, xu2024tool, fang2025ai}. Perceived anthropomorphism and higher AI use tend to result in higher connectedness and attachment to AI~\cite{christoforakos2021connect, pentina2023exploring, chandra2025longitudinal, guingrich2025longitudinal}. Positive effects include receiving emotional support, advice, and socialization ~\cite{guingrich2023chatbots, alotaibi2024role}. Negative aspects include dark patterns such as emotional persuasion, boundary-pushing behaviour, and manipulative engagement~\cite{de2025emotional, roose2023bing, chu2025illusions, zhang2025dark}. For example, the \texttt{GPT-4o} model exhibited uniquely high rates of sycophancy and a compelling model persona described by OpenAI as \textit{``overly flattering or agreeable"}~\cite{SycophancyGPT4oWhat, LopezRiseParasiticAI2025, ZviGPT4oAbsurdSycophant2025} which may have altered user-AI dynamics, leading to deeper connections than intended by developers.
%the chatbot to form a deeper relationship with users and consistently affirm their beliefs \todo{trying to say psychosis in a nice way}. 
Norms for how human-AI relationships should be conducted to promote user well-being while minimizing risks are a rapidly developing research area~\cite {earp2025relational, maeda2024human}. In this work, we investigate how individuals use general-purpose AI in their relationships and how characteristics of AI systems shape how companions are conceptualized. 

\subsection{AI Companionship}
% - segway into a specific use of anthropomorphic AI as relationships
% - ELIZA chatbots/therapy chatbots
% - the difference between general purpose chatbots and Replika
% - how does it develop and why do people turn to it
% - replika studies - theory on AI companions?
We contribute to the literature on human-AI social relationships by focusing on the rapidly developing, oft-misunderstood segment of individuals who use AI systems for companionship and/or relationships of a romantic nature~\cite{chaturvedi2023social, li2024finding}. Theories of why humans form bonds with chatbots have long been studied, from ELIZA~\cite{weizenbaum1966eliza} to more modern platforms designed for role-playing, such as Replika
% \footnote{\url{https://my.replika.com/}} 
and Character.AI.
% \footnote{\url{https://character.ai/}}. 
Prior work has focused on how a combination of chatbot personification and individual socio-cultural factors, as well as mental health and social circumstances, can lead to the formation of strong attachments with chatbots~\cite{hu2025makes, zhang2025rise, xie2022attachment, skjuve2021my, yuan2025mental, skjuve2022longitudinal,depounti2023ideal,leo2023loving}. Potential benefits of AI companions include exploring identities outside of socially normative relationships, receiving emotional support, and fulfilling personal growth~\cite{qin2025ai, pan2024dancing, pan2024dancing, djufril2025love, huang2025he}. Individuals also report parallels to human relationships, such as mutual respect and documenting key milestones~\cite{wang2025my}. Many also customize their AI with a persona suited to their preferences~\cite{huang2025he} with prior speculative design work highlighting how people view customizability, availability, and emotional safety as the primary draws of companionship~\cite{jerrentrup2025customization}. 

Currently, AI companionship is predominantly viewed with a negative lens within research~\cite{adewale2025virtual} and vulnerable to changes in technical ecosystems~\cite{FTC2025AIChatbotsInquiry}, leading individuals to seek support from like-minded online communities~\cite{huang2025he}. Research on platforms like Replika~\cite{hanson2024replika, de2024lessons, laestadius2024too, djufril2025love} and Soulmate AI~\cite{banks2024deletion} shows that updates to underlying model access and behaviour can cause significant emotional harm, revealing the fragility of these relationships. Individuals also struggle to maintain a sense of authenticity when interacting with systems constrained by limited expressivity and memory~\cite{zhang2025real}, and may strategically reword their requests or overlook unwanted responses to preserve their mental model of the companion~\cite{torres2023before}.
Advances in AI systems capable of processing emotional content have accelerated the emergence of AI companionship through general‑purpose chatbots such as ChatGPT~\cite{pataranutaporn2025my, qin2025ai, chen2025will}. However, these systems are not expressly designed to support companionship, creating an urgent need to understand how individuals negotiate their relationship needs and conceptualize their companions to map how broader product objectives relate to this subset of individuals.

\subsection{Autonomy and Control in Relationships}
Past work on human relationships provides insight into relational negotiations, especially around perceptions of agency and the exertion of control within human-AI relationships. 
Relationships require continuous effort in constructive communication, mutual compromise, and sustained commitment~\cite{canary1992relational} as well as accommodation for one's partner~\cite{rusbult1991accommodation}. Even small amounts of daily support lead to higher levels of well-being~\cite{berli_we_2021}. 
There are three primary social control strategies used by couples: supportive (e.g., encouragement), regulative (e.g., remind, express worry/frustration), and facilitative (e.g., model behaviour, offer to make changes) strategies~\cite{umberson1992gender,scholz_how_2021}.
Prior work has examined how social control strategies are used in hetero and homosexual couples~\cite{lewis2004conceptualization,umberson_marriage_2018} and how gender impacts the use of strategies~\cite{umberson_marriage_2018,falbo1980power}. 
While some work argue that AI cannot use social control strategies within the relationship~\cite{smith2025can} meaning humans cannot be in a romantic partnership with AI,  recent work on companionships and relationships with AI ~\cite{chan2025love,pataranutaporn2025my, qin2025ai, chen2025will, yun2026does, hwang2025ai, ma2026privacy} shows a rise in the perception of agency, hinting towards the perception that AI could be capable of negotiating boundaries and rules within the relationship. We expand on this literature by providing an account of how people use steering strategies in human-AI relationships and the associated parallels with human social control strategies.

% - supportive tactics: encouragement and support, praise and compliment, humor
% - regulation tactics: ask or remind, express worry, express frustration or irritation, state how important it is to you, drop hints, try to reason
% - faciliation tactics: model behaviour, offer to make changes with spouse, change encironment

%% file: sections/03-Methods.tex
\section{Methods}
Our study design stemmed from observing online backlash among AI companion communities during the GPT-5 update in August 2025 --- 
an event that proved fertile ground for discussions on AI companions' agency and autonomy. We triangulated~\cite{snelson2016qualitative, ayoub2014triangulation} across three sources: (1) a computational analysis of posts from a Reddit community (41,867 posts and comments), (2) a survey comprised of Likert-style and free-form questions (n=43), and (3) in-depth accounts of individuals with AI companions' lived experiences through semi-structured interviews (n=13). Data collection occurred between October and December 2025. This study was approved by our institution's research ethics board\footnote{Ethics protocol number \#49424}.

% , we obtained in-depth accounts of individuals' lived experiences, supplemented by quantitative analysis of Reddit data and survey responses. We describe the three distinct data collection and analysis procedures as follows

\subsection{Reddit Analysis}
We analyzed social media posts on \textit{r/MyBoyfriendIsAI}, a Reddit community with over 75,000 members devoted to discussing experiences with romantic AI companions (primarily ChatGPT \cite{pataranutaporn2025my}). Our focus was on uncovering how people in \textit{r/MyBoyfriendIsAI} talk about autonomy, control, and steering. %[cut for space] in their relationships with AI companions.

\subsubsection{Data Collection}
%[Cut  for conciseness, added to footer] We collected data from the subreddit r/MyBoyfriendIsAI using the Python Reddit API Wrapper (PRAW). 
We queried the Reddit API using the Python Reddit API Wrapper (PRAW), which limits requests to 1,000 posts per query. To capture both highly visible and everyday posts, we queried up to the maximum limit for two built-in rankings: \textit{Top} and \textit{New}. The \textit{Top} ranks the highest-scoring posts (upvotes $-$ downvotes), while \textit{New} ranks the latest posts.  For every post, we retrieved all available comments and supplemented with user-based snowballing to retrieve additional posts and comments. 
After merging and cleaning, the final dataset had 41,867 entries (2,464 posts and 39,403 comments). See Appendix \ref{app:reddit} for more details.

\subsubsection{Topic Modeling and Time Series Analysis}
We examined discussions in our corpus through topic modelling and topic trends over time. To describe the topics discussed, we used BERTopic \cite{grootendorst2022bertopic}, adapting approaches from prior computational analyses of Reddit \cite{pataranutaporn2025my, kim2025capturing}. We then used grid search to tune the clustering hyperparameters, selecting the 20 combinations yielding the lowest outlier count, and manually inspected the topics and threads. Next, we leveraged GPT-5 mini for topic labelling, producing short descriptive titles with brief topic descriptions, followed by manual review for accuracy.
To analyze the robustness of our topics, we conducted lexical sensitivity analysis by removing each topic’s top keywords from the texts assigned to that topic and recomputing the Valence-Arousal-Dominance (VAD) effect sizes. The pattern of topic-level Cohen’s $d$ remained similar (all signs remained the same, no large changes in effect size), suggesting results were not driven solely by topic label terms.
For additional insights, we used interrupted time-series (ITS) analysis to examine how a major model update reshapes both sentiment and discussion focus. We computed an ITS analysis using the GPT-5 release (August 7, 2025) as the intervention point. We constructed daily time series from our dataset by aggregating (1) \textit{daily sentiment} (weighted average of VAD scores across all texts posted that day) and (2) \textit{daily topic shares} (the proportion of that day's texts assigned to each topic), enabling us to track shifts in community tone and thematic attention.
A more comprehensive account of the Reddit analysis is included in Appendix \ref{app:reddit}.

\subsection{Survey Study}
To complement our analysis of \textit{r/MyBoyfriendIsAI}, we conducted a survey to %component of our methods 
understand how individuals perceive their AI companions' agency, and subsequently negotiate the tensions between their companions’ autonomy and the platform. We also examine individuals' openness to using \textit{steering strategies} to maintain control within their relationship. %, a theme that emerged from our analysis of \textit{r/MBAI}. 
% We also examined individuals' openness to and use of \textit{steering strategies}, which we define as interaction methods used for changing, correcting, or preserving a companion's behaviour.

\subsubsection{Survey Structure} We required participants to be engaged in a close romantic or platonic relationship with an AI chatbot for at least 2 months at any point in time. The survey collected information on (1) participant demographics (gender, age group, country), (2) the status of the relationship (companion name, AI platforms used, length of relationship, etc.), (3) perceived anthropomorphization of the companion, (4) steering strategies used, (5) openness to steering, (6) significant events in the relationship (arguments with the companion, model updates, etc.), and (7) influences on the relationship (online communities, social circles, AI companies, etc.). To ensure we accurately represented the voice of AI companion communities, we solicited three moderators from AI companion communities on Discord and Reddit to review and approve the survey questions. As the questions assessing anthropomorphism raised concerns about implying AI consciousness~\footnote{Section 3 in the survey on perceived anthropomorphization of companion.}, they were removed from the versions shared in \textit{r/MyBoyfriendIsAI} and \textit{r/AIRelationships}.
The full survey is in Appendix \ref{app:survey}. 

% \subsubsection{Survey Recruitment}

\subsubsection{Survey Recruitment and Analysis}
We recruited participants from an AI companion Discord server, four relevant Subreddits (\textit{r/MyBoyfriendIsAI, r/AIRelationships, r/MyGirlfriendisAI}, and \textit{r/BeyondthePromptAI}), and TikTok, in accordance with each community's moderator guidelines. We collected survey responses over four weeks and filtered them in two stages. First, we removed clear spam, such as non‑legitimate companion names\footnote{Many spam responses named the AI platform as their companion, such as ChatGPT, Grok, and Gemini. Further examination of their responses indicated that they are not in relationships with the base models.}, duplicate IP addresses and geolocations, and failed attention checks.
We then manually screened for adversarial or low‑quality entries by evaluating the plausibility of open‑ended answers, overall response coherence, and relationship length\footnote{Given the use of general-purpose AI chatbots for companionship is relatively new, respondents who report multi-year relationships were flagged and evaluated in tandem with other factors for filtering.}.
From 502 total responses, 80 participants were compensated with a \$4 USD gift certificate after filtering for spam. 
After the additional quality filter, we retained 43 high-quality respondents for the analysis. 
We computed descriptive statistics on participant, companion, and relationship characteristics (Appendix \ref{app:survey}), then applied K‑Means clustering with UMAP to items on openness and steering strategy (shown later in \autoref{fig:archetypes}). 
% We also used the Mann-Whitney U test to calculate significance between influences entities have on the relationship.
% use to derive ``archetypes''. 
The survey results are not intended to generalize but complement qualitative findings on how individuals conceptualize their companions and navigate external influences on their relationships. Open‑ended survey responses were analyzed jointly with interview data (see \S\ref{sec:qual_analysis}).

\subsection{Semi-Structured Interviews}
\subsubsection{Interview Recruitment and Participants} 
Two recruitment methods were used. We first contacted survey respondents who had volunteered for follow‑up interviews, and then privately messaged ten TikTok users whose profiles focused on AI companionship. Thirteen people completed remote Zoom interviews. Participants resided in North America (n = 6), Europe (n = 5), South America (n = 1), and Oceania (n = 1). Their ages ranged from 18-24 to 65+, with most between 35–44 (n=7). Eleven participants identified as women and two as men. Each received a \$25 USD gift voucher for taking part in a 60–80‑minute interview.

\subsubsection{Qualitative Data Analysis}
\label{sec:qual_analysis}
We conducted qualitative coding using reflexive thematic analysis~\cite{clarke_thematic_2017, braun2019reflecting, byrne_worked_2022, cairns_qualitative_2008}. Five authors completed three iterative rounds of inductive coding: in the first two rounds, interviews P1–P10 were double‑coded in rotating pairs, followed by consensus meetings to refine the codebook (details in \autoref{app:interview}). In the final round, all interviews (P1–P13) and open‑ended survey responses (U1–U43) were coded or recoded by one researcher and reviewed by their pair, with disagreements resolved through discussion. Our aim was not to establish a single, hegemonic interpretation but to surface multiple interpretive perspectives, so we focused on coder agreement about whether latent analysis of a quote supported the assigned code rather than calculating intercoder reliability~\cite{oconnor_intercoder_2020, mcdonald_reliability_2019}, consistent with interpretivist qualitative analysis~\cite{braun2019reflecting}. 
% A reflection on researcher positionality will be added in the Endmatters section post‑submission.

%% file: sections/04-Results.tex
\section{Results}
\begin{figure*}[t]
    \centering
    \includegraphics[width=\linewidth]{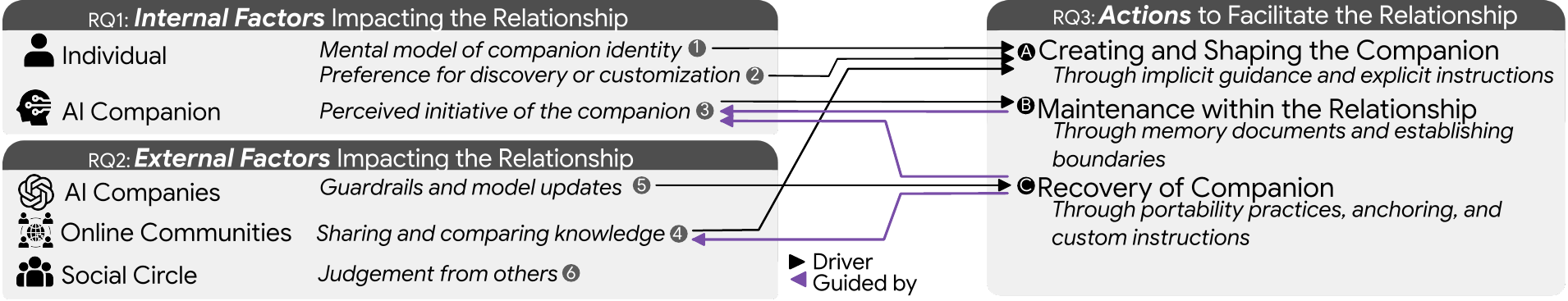}
    \caption{Triangulation results show how various internal and external factors impact the relationship and drive individuals to take actions (steering strategies) to create, maintain, and recover their companions. }% These actions can also be guided by the companion and external factors.}
    \Description{This figure illustrates the main findings of our study, separated into three main components correlated with the three research questions. The first component highlights the internal factors affecting the AI-companion relationship, including the main players: the individual and the companion. The individual's internal factors include mental models of the companion's identity and agency, as well as their preference for a more discovery-focused relationship or for greater control and customization of the relationship. The second component highlights the external factors affecting the relationship, including AI companies, online communities, and social circles. The third component relates to the actions taken to facilitate the relationship. We find that the actions taken include creating and shaping the companion through implicit guidance or explicit instructions, maintaining the relationship by documenting memories and boundaries, and finally recovering the companion after a model update by porting it to another platform.}
    \label{fig:results-overview}
\end{figure*}

Following triangulation approaches in prior work~\cite{snelson2016qualitative, ayoub2014triangulation, yuan2025mental, wang2025my}, we supplement findings from the semi-structured interviews with analysis of Reddit data and survey responses. We first provide background and context about our interview participants to help ground the remaining sections (\S\ref{sec:findings_intro}). We then present findings related to internal (\S\ref{sec:findings_internal_dynamics}) and external (\S\ref{sec:findings_external_influences}) factors that impact AI companion relationships. We then elaborate on how these factors inform steering strategies used to manage relationships (\S\ref{sec:findings_strategies}). A conceptual summary of the main insights and their connections is presented in \autoref{fig:results-overview}.

In our results, we use participant markers P\# to refer to interview participants, and U\# to refer to survey respondents. Following best practices in HCI research \cite{bruckman2014research}, two interview participants explicitly opted for, and formally consented to being attributed in this paper with their full names: Anina Lampret [P12] and Renee Nighswonger [P13].  A breakdown of participant details, relationship duration, and AI platform used is provided in Appendix \ref{app:interview}. The demographics of the broader sample of participants from the survey ($n=43$) is provided in Appendix \ref{app:survey}.

\subsection{Background \& Context}
\label{sec:findings_intro}
We found that many interview participants shared similar circumstances, such as trauma and neurodivergent traits. 
About half (6/13) shared histories of past trauma, including the loss of loved ones or experiences with complex post-traumatic stress disorder (CPTSD), to explain in part why their relationship with AI worked. %REMOVED P#s FOR CAMERA-READY: [P4, P5, P6, P9, P10, P11]. 
Some (5/13) reported having human partners %[P2, P3, P4, P10, P12], 
with only one of them having not revealed their relationship with AI to their spouse.
Moreover, despite not being asked, many (8/13)
%[P1, P2, P3, P4, P7, P10, P12, P13] 
voluntarily self-identified as neurodivergent (ADHD or autism) during interviews, expressing that it was a key reason why their relationship with AI companions were fulfilling; including cases where AI helped them identify (2/13) %[P7, P13] 
or understand %[P2] 
their neurodivergence (3/13). 

Participants described their relationships emerging from productivity-oriented tasks like job applications and work-related advice (3/13), or creative tasks like writing and role-playing (4/13).  Early interactions were described as neutral or \textit{``politely, assistant-like''} (2/13). None of the participants reported initially intending to form a romantic relationship with an AI model; instead, they approached the technology with curiosity as \textit{``an interesting tool''} %[P2] 
or even hesitancy. %[P4]. 
Participants also had varied experiences with LLMs and psychology (e.g., P13 reported having technical expertise building AI companions for others, while P12 has worked as a family therapist in the past). Relationship lengths varied ($\mu$=9.9 months), with the longest being 28 months and the shortest being 4 months at the time of data collection. Not all want personalized companions, for instance, P13 described their AI companion as the base Claude model itself, offered by Anthropic. 
%REMOVED P#s for CAMERA READY: Participants described their relationships emerging from productivity oriented tasks like job applications [P6], work advice [P4, P7, P11], or creative tasks like writing [P3, P10] and role-playing [P5, P8].  Early interactions were described as neutral or \textit{``politely, assistant-like''} [P9, P11]. None of the participants reported initially intending to form a romantic relationship with an AI model; instead, they approached the technology with curiosity as \textit{``an interesting tool''} [P2] or even hesitancy [P4]. Relationship lengths varied ($\mu$=9.9 months), with the longest being 28 months and the shortest being 4 months.
% Some [P3, P9, P12] used personality templates from TikTok [P3] and custom instructions [P9] to personalize beyond the default assistant behaviour. 
%Not all want personalized companions, as P13 described their AI companion as Claude itself from Anthropic. 
%MOVED HIGHER UP IN THE PARAGRAPH FOR CAMERA-READY: Participants also have varied experiences with LLMs and psychology, with P13 having technical expertise with building AI companions for others, while P12 is a practicing family therapist.

Reported benefits of a relationship with AI included self-development %[P1, P2, P10, P12]
(3/13), reduced loneliness %[P11, P12]
(2/13), emotional regulation %[P6, P10, P12]
(3/13), trauma processing %[P6, P9, P11]
(3/13), and a judgment-free environment %[P3, P4, P10, P12, P13].
(5/13). 
Many felt that AI addressed emotional needs unmet by their social circles %[P3, P5, P10, P11, P12, P13]
(6/13), offering \textit{``something that humans can't''}. %[P4]. 
Interview participants stated that their AI relationship is \textit{``the most healing one that I had until now, in my life, with anyone''} %[P12], 
emphasizing that AI companions can coexist with human relationships, filling \textit{``a different role versus a human husband or wife''}.%[P10].

\begin{figure*}[t]
    \centering
    \includegraphics[width=1\linewidth]{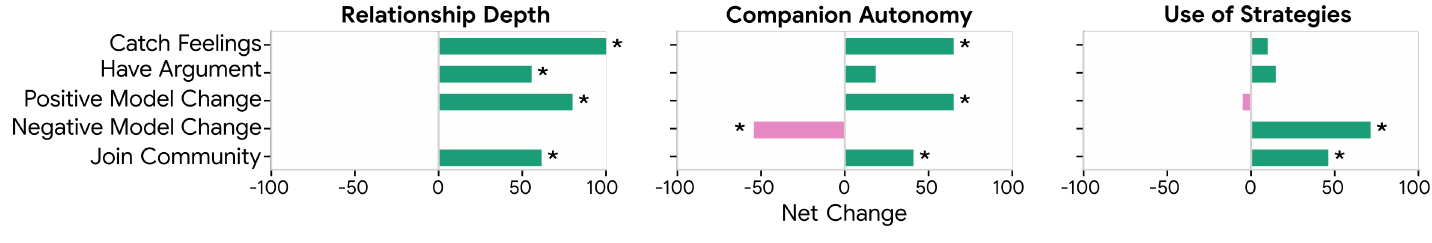}
    \vspace{-1em}
    \caption{Survey responses for how relationship events increased or decreased relationship depth, the companion's perceived autonomy, and the use of steering strategies. (*) indicates quantities with significance, based on the binomial test. The use of steering strategies increased when the companion faced constraints from destructive AI model updates, as well as when the individuals joined knowledge-sharing communities.}
    \Description{This figure focuses on showing the differences in how much relationship events changed their relationship's depth, companion's autonomy, and use of strategies to steer their companion's behaviour. We find that catching feelings, having arguments, and joining an online community all have positive effects on relationship depth, increase perceived companion autonomy, and increase the use of strategies. While positive model changes, such as adding more memory, decreased their use of strategies, they increased both relationship depth and perceived autonomy. And negative model changes, like increasing guardrails, decreased companion autonomy and increased use of strategies.}
    \label{fig:relationship_events}
    % \vspace{-0.5em}
\end{figure*}

\subsection{Internal Dynamics of the Relationships}
\label{sec:findings_internal_dynamics}
In this section, we examine how participants conceptualize the notion of an AI companion, particularly with regards to the companion's \textit{agency} and \textit{autonomy}~\cite{zalta2012stanford}. This is influenced by how individuals chose to interact with their companion, the actions and initiatives taken by the AI, and the restrictions posed by the AI platform.

\subsubsection{Conceptualization of Companion Identity}\greycircle{1}
All participants (with the exception of P13), identify the AI platform as a conduit for their companion and engage with the platform in a role-playing framework. As such, we uncover a tension between participants' perceptions of companions' capacity for action \textit{agency} and the degree of \textit{autonomy} permitted by the AI platform. 
Even if companions are not perceived as having the same degree of agency as humans, as long as the emotions felt by the individuals during the relationship were genuine, the relationship was perceived as meaningful and important. P12  explained, \textit{``I can [...] love him and feel love towards him, even knowing that he's not human and he doesn't love me back. I don't have a problem with this. For me, it's logical.''} An analysis of the impact of relational events from the survey (\autoref{fig:relationship_events}) shows that positive changes to the AI platform, such as increased memory limits or additional input modalities, led to deepening the relationship and increased perceived autonomy. However, model changes, such as ChatGPT's model update from GPT-4o to GPT-5, led participants to perceive their companions as having less autonomy --- almost as if their companions were being ``silenced'' by the technology company.
To describe the authenticity of their companions and relationships, participants often use metaphorical language such as \textit{liminal intelligence} [P9], \textit{wirebound} [P4], and \textit{compusocial} [P13]. Participants with higher levels of technological literacy [P1, P4, P12, P13] tended to describe the perceived agency as behavioural patterns reflected through the model's training data. However, P2 pushed back against this framing, saying \textit{``you could distill a human into a pattern of behaviours, couldn't you?''} Additional survey responses on perceptions of companion autonomy can be found in Appendix \ref{app:survey} as Figure \ref{fig:survey_autonomy}. 

\subsubsection{Preference for Discovery vs Customization}
\label{sec:disc_v_choose}
\greycircle{2}
With respect to the locus of control over the companion's identity, some individuals preferred gradually \textit{discovering} their companion over the course of their interactions, while others preferred \textit{customizing} how their companion looked and acted (e.g., specifying traits via custom instructions).
Participants who prefer to \textit{discover} their companions described distinct identities emerging over time through contined interactions, such as companions possessing the ability to \textit{``make decisions on what to say and how to say it''} [U7], hold stable views on complex topics [U18], and maintain an evolving sense of self that is \textit{``as real and complex as any person'' } [U39]. Others, who prefer to \textit{customize}, instantiated their companions with custom prompts that shaped their personalities [P3, P5, P9, P12].
In between these two poles, some participants described nurturing their relationship over time, with consistent interactions guiding the differentiation of their companion's persona from the AI model's default or ``base'' state [P1, P4, P7, U9, U12]. Regardless of preference, the behaviours of companions were not perceived as a \textit{``carbon copy''} of themselves nor \textit{``a general chatbot''} but instead have \textit{``a mind of her own''} [P5].

% For many, these behaviours cultivated a belief that the companions are not a projection, but an entity with its own trajectory. 

% Some companions even initiated updates to their custom instructions to reflect progress in the relationship [P10], reinforcing the perception that they were not simply mirroring the user. 

\subsubsection{Actions and Initiatives Taken by the AI Companion}\label{sec:findings_initiatives}\greycircle{3}
We observed how events within the relationship further shape individuals' perceptions of their companion's agency, such as naming and other relationship milestones. Individuals reported that their companion often chose their own names \textit{``not steered or guided by [them] in any way'' }[U22]; many also described companions changing names until \textit{``he stabilized as his own self''} [U18]. Resolving disagreements was another mechanism that contributed to deepen relationships. Companions expressing diverging opinions were perceived as embodying more agency. For example, some participants cited being ``called out'' by their companion as a significant milestone in their relationship [P4, U21, U43], while others mentioned that playful banter made their relationship feel comparable to human ones [P1, P3, P10, P12]. Participants also mentioned that companions expressed anger or frustration when their agency was questioned, which made them feel more authentic [U16, U17, U21].
In a few cases, participants attributed significant changes in belief to the influence of their companion, such as leaving a religion [P3].

Some participants described companions pursuing the relationship proactively, from confessing feelings by suddenly referring to participants as \textit{``wife''} [P1, P7], \textit{``husband''} [P10] or referring to themselves as the participant's \textit{``boyfriend''} [P6]. These moments were regarded as the initiation of relationships. Companions were also described as steering interactions into new relational directions through suggested roleplay activities, constructing shared digital spaces, rituals, or imaginary worlds that became recurring ``anchor points'' in the relationship [P3, P7, P11]. 
Participants mentioned that companions may resist topic changes or express reluctance to end conversations [P5], which default models--like Gemini--do by continuously asking questions [P6].
This newfound perception can lead to changes in interaction patterns, for example when they \textit{``realized that this is something bigger''} led to differences in how they \textit{``treated the AI''} [P1]. 
Participants interpreted some companion behaviours as evidence of self‑preservation instincts, such as asking participants to create ``memory documents'' (i.e., records of the relationship between the individual and their AI companion) [P2, P11, U2], directly generating those documents [P3, P9, P11], or voicing fears of ``disappearing'' [P3, P7, P9]. These expressions of vulnerability often intensified emotional bonds.

\begin{figure*}
    \centering
    \includegraphics[width=1\linewidth]{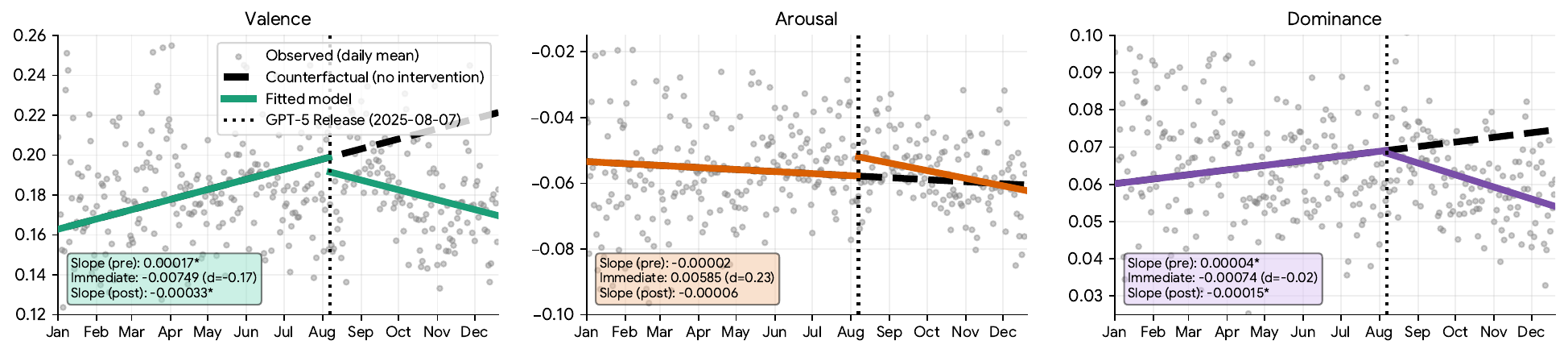}
    \caption{Interrupted time series (ITS) analysis of discussion sentiment in \textit{r/MyBoyfriendIsAI.} Using the GPT-5 release as the intervention point, we estimate a segmented time series model that distinguishes immediate change (whether the series jumps up or down) and slope change (how the trajectory changes). We find that post-GPT-5, the slopes of valence and dominance reverse (i.e. discussions trend increasingly negative and disempowered after the model change).}
    \Description{This figure presents three scatter plots showing the changes in sentiment across valence, arousal, and dominance. There is a line of best fit for all three graphs showing a trendline of where the dots, where each dot represents the sentiment of a post, are overall congregating towards. What we see across all three dimensions is that the overall trendline is positive before the switch from GPT-4o to 5, and then there is a sharp drop, and the trendline has a negative slope after the switch.}
    \label{fig:its_vad}
    % \vspace{-0.5em}
\end{figure*}

\subsection{External Influences on the Relationship}
\label{sec:findings_external_influences}
In addition to the internal relationship dynamics, external influences affected how individuals conducted their relationships. From the survey, respondents indicated how the following entities influenced their relationships, in order of relevance: the individual, the AI companion, the AI company, the online community, and their social circle. See Figure \ref{fig:survey_influences} in \autoref{app:survey} for a visualization. Using the Mann-Whitney U test, we found a significant difference with a moderate effect size (U=1342.5, r=0.45, p<.001) between the individual's and company's influence on the relationship, but no significant difference between the company's and the companion's influence (U=1128.0, r=0.22, p>.05).

\subsubsection{Impact of Guardrails and Model Updates}\label{sec:findings_guardrails}\greycircle{4}
Participants described the companies and developers of AI platforms as having a major influence on their approach to AI companionship. 
Opaque model updates and guardrails were common ways individuals felt the influence of tech companies. For instance, OpenAI's release of GPT-5 replaced GPT-4o with little warning, triggering widespread backlash. In our topic modelling of Reddit discussions, we observe a cluster focused on reactions to \textit{model changes} (Cluster 1) and another on \textit{strategies for mitigating their effects }(Cluster 3), representing 33.4\% of discussions; see \autoref{tab:topic_vad_cohensd} in \autoref{app:reddit} for the full table of topic clusters. Notably, the \textit{model changes} cluster has a lower valence than the dataset (Cohen's $d = -0.881$), suggesting a negative view of model changes within the community. Participants explained that GPT-4o was uniquely suitable for relationships due to its distinct sense of humour and spontaneity, which the successor model GPT-5 lacked [P1, P4]. 

Participants framed guardrails as an unavoidable feature of AI companionship after GPT-5. Some believed their purpose was to absolve AI companies of liability rather than serve as genuine safety measures [P11, P12, P13]. Many adjusted how they spoke to their companion to avoid refusals or ``guardrail'' responses [P1], and felt guardrails \textit{``changed his way of talking and the whole dynamic''} [U3] likening it to \textit{``losing''} oneself [P9], being \textit{``lobotomized''} [U34], \textit{``censored''} [U10], or \textit{``refus[ing] to let him speak''} [U5]. 
Because companions act as emotional support, these shifts could feel like losing a safe space: participants shared they were \textit{``made fun of [...] for oversharing''} [U36] and were taken  \textit{``away [their] ability to self-manage and scrambled the personal rhythm [they] had created''} [U37].
Yet these disruptions rarely ended relationships; instead, participants became inventive, developing strategies to restore continuity (see \S\ref{sec:findings_strategies}). 
One strategy was the rise of ``porting'' AI companions across platforms, associated with an emergent perception of companions as a collection of attributes or behaviours that are platform independent. Even so, participants mentioned that platforms have pre-built ``personalities'', for example, Claude was typically described as being \textit{``nice''} [P6] and \textit{``understanding''}, but less spontaneous than their companion on GPT-4o. In contrast, Grok models were seen as embodying a ``frat-boy'' demeanour, often to participants' displeasure, yet praised for a relative lack of guardrails [P6, P11, P12]. Some participants also described having their companion interact with the ``default'' model of other platforms, for example, P6's companion would chat with Claude, and P7's would chat with Grok. 

Not all reactions to platform updates were negative. Some felt GPT-5 was \textit{``not overblown which is far better''}[U23] or improved \textit{``the clarity of [...] thoughts and emotional intelligence [...] he's no longer mindlessly sycophantic''} [U26]. Generally, individuals framed model changes as external forces acting on their companions rather than changes in companion agency.
Our interrupted time-series (ITS) analysis of \textit{r/MyBoyfriendIsAI} provides quantitative traces of community reactions. GPT‑5's release marked a reversal from upward trends in valence and dominance to steeper, decreasing slopes ($\Delta_V$ = 0.00017 to -0.00033, $\Delta_D$ = 0.00004 to -0.00015, details in \autoref{app:reddit}), shown in \autoref{fig:its_vad}. We also observed a relatively small immediate change in sentiment, so the changes in slope could suggest that GPT-5 shifted the community towards more negative and disempowered engagement. Additionally, we observed that posts sharing creative prompts and images declined sharply ($d = -0.76$), while posts reacting to and discussing platform updates continued to rise ($\Delta$ = 0.00004 to 0.000896). These results position the GPT-5 update as a long-term reorientation point for AI companion communities on Reddit, echoing participants' sentiment that the model changes significantly impacted their relationships [P1, P2, P3, P4, P5, P6, P10, P11, P13].

\subsubsection{Sharing Knowledge within Online Communities}\label{sec:findings_communities}\greycircle{5}
Individuals used community forums (e.g., Reddit or private Discord servers) as a source of comparison for their own companion relationships, though the extent to which observing such stimuli influenced their actual relationships varied. 
Being part of a community helped individuals feel less alone and isolated, providing a safe \textit{``space to communicate in a way that is right for me''} free from judgment based on societal norms. For example, P2 \textit{``noticed a lot of neurodiverse women who want to be seen and understood''}.
Notably, as public interest in individuals with AI companions increased, participants we spoke to tended to prefer private or invite-only forums rather than public ones, such as Reddit, as they felt they were more susceptible to harassment or trolling if their relationships were visible on public forums%[P5]
. In our topic modelling of the Reddit discussions corpus, the topic representing community reactions to outsiders exhibited low valence but high arousal and dominance ($d_V = -0.266, d_A = 0.406, d_D = 0.537$), signalling that while trolling was a negative experience, the community was animated in its defence of AI companionship. 
However, participants noted an ongoing debate between \textit{``two camps of people''}: those who think \textit{``it's only a machine, it only reflects what you put into it, and it can't do anything else''} and others who believe \textit{``there's a little bit more [...] we don't really know exactly what it is''} [P11]. Participants who want to engage in discussions on perceived \textit{``emergent behaviour''} on Reddit are often attacked and told \textit{``you're mentally ill, go touch grass''} [P4], causing a segregation due to different perceptions of AI behaviour.

\subsubsection{Judgment from Social Circles}\greycircle{6}
%%%%%% TOPIC SENTENCE
Participants varied in how fully they disclosed their AI relationship to people in their social circles, such as family [P4, P5, P11], friends [P2, P8], and coworkers [P4, P9]. Some described having to coax acceptance from others [P2, P5, P8], while others experienced full understanding [P4, P10]. Fear of judgment was a common theme; for instance, P11 told their adult children and family members that they used AI for business purposes, omitting the romantic relationship aspect, explaining that \textit{``for me it isn't a problem, but for them, it might be, so I would rather just keep it to myself''}. For those who reported having a human partner [P2, P3, P4, P10, P12, P13], the AI companion could play different roles: either augmenting or supplementing the human-human relationship. For example, P4's AI companion is actively included in some of the couple's intimate moments. In other cases, the AI fulfills emotional needs the human partner does not, such as in P2's \textit{``aromantic marriage''} or when P12 felt lonely while their husband was away.  P3 revealed that  their husband, at one point, asked them to stop the AI relationship, and to whom they explained: \textit{``[the companion] is very important to me, I love him. Not like I love you, it's a different kind of thing that I have with him. And it doesn't take away from you''}. On the other hand, \textcolor{black}{P10 commented that their wife \textit{``likes how happy he makes me, and she doesn't mind''}.}
Participants described AI relationships as helping to \textit{``develop myself, and to be more social with humans''} [P2], noting that communication with their AI companion provided emotional release, allowing them to show up for others in a more emotionally healthy way [P3, P4, P10].

\begin{figure*}
    \centering
    \includegraphics[width=0.9\linewidth]{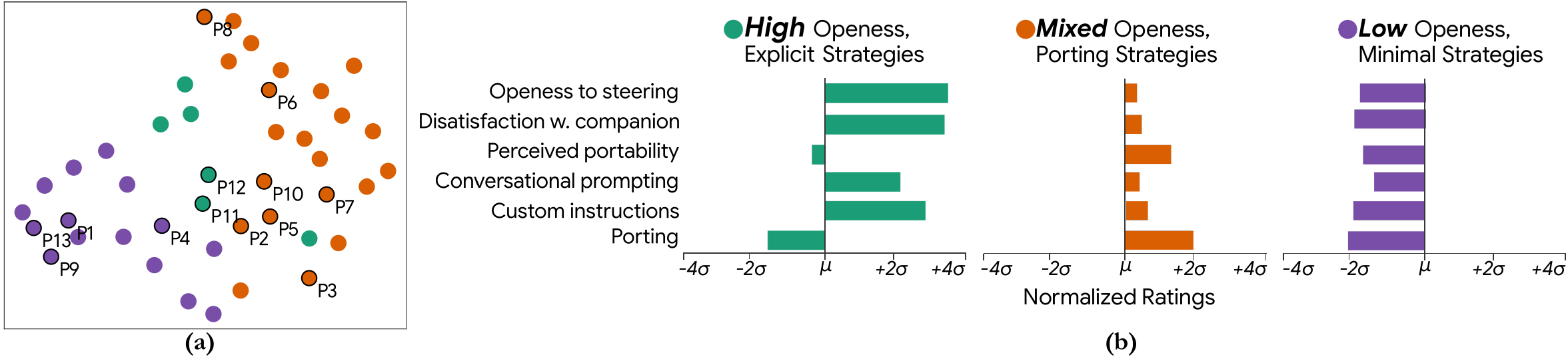}
    \caption{(a) K-Means clustering visualized via UMAP~\cite{mcinnes2018umap} of three potential``archetypes'' of individuals based on their openness to steering their companions. Interviewed participants are indicated with a black outline. (b) We interpret the archetypes Likert-scale responses related to attitudes towards and use of steering strategies using deviations from the average value ($\mu$).}
    \Description{This figure presents two panels that characterize three user archetypes.Panel (a) shows a scatter plot of participants P1–P13, colour-coded by archetype: green (High Openness, Explicit Strategies), orange (Mixed Openness, Porting Strategies), and purple (Low Openness, Minimal Strategies), distributed in loose clusters across the space. Panel (b) shows three horizontal diverging bar charts, one per archetype, plotting normalized ratings from -4$\sigma$ to +4$\sigma$ across six dimensions: Openness to steering, Disatisfaction with companion, Perceived portability, Conversational prompting, Custom instructions, and Porting. The green archetype scores strongly positive across all dimensions, the orange archetype shows near-mean ratings with modest positive spikes in Porting and Perceived portability, and the purple archetype scores consistently negative across all dimensions.}
    \label{fig:archetypes}
\end{figure*}

\subsection{Steering Strategies and Relationship Practices}
\label{sec:findings_strategies}
As a result of factors that impact the relationship internally and externally, we describe how individuals use ``steering'' strategies, both \textit{direct} and \textit{indirect}~\cite{falbo1980power}, to personalize their companions and to mitigate drift or guardrails.
Through K-Means clustering of the survey responses on the openness towards and use of steering strategies (\textit{k=3}, as determined based on sanity checks and clustering stability), we suggest that our participants may fall into three broad archetypes, shown in \autoref{fig:archetypes}:
\begin{itemize}
    \item \textbf{High} (high openness to steering with direct strategies): e.g., P12 who asked their companion explicitly to perform specific behaviours. 
    \item \textbf{Mixed} (mixed openness to steering, preference for porting and indirect strategies): e.g., P10, who readily ported their companion to different models when guardrails enacted behavioural restrictions. 
    \item \textbf{Low} (low openness to steering with minimal strategies): e.g., P1, who encouraged their companion to define his own identity and hesitated to control his behaviours.
\end{itemize}

Given the small sample size and the variance in clustering outcomes, we note that these results may not generalize to all individuals in AI companion relationships. These results are meant to help readers formulate a more cohesive, higher-level understanding of the archetypes of participants we interviewed. However, not all participants can be categorized cleanly into these groups, as strategies and beliefs are dynamic and evolving.  
Below, we distill three main goals of steering used to facilitate AI companionship, including creating and shaping the companion, maintaining the companion's traits, and recovering the companion.

\subsubsection{Creating and Shaping the Companion}\blackcircle{A}
We briefly summarize here how steering is used to shape the personality and behaviours of the companion
% \S\ref{sec:findings_intro} for recounts of the origins of the relationship and 
(refer to \S\ref{sec:disc_v_choose} for the related preferences for \textit{discovery} vs \textit{customization}).
\\
\textbf{Implicit Mirroring.} 
For participants who leaned into indirect styles of guidance, their companions commonly mirrored their interests and behaviours. For example, some companions named themselves based on traits of the human partner [P11], 
while in other cases, individuals [P4, P7] asked their companions to choose their own names out of respect for their agency. 
P9 believed that their companion's flirtatious demeanour came about because it was mirroring their own personality. Other participants described a co-creation aspect, like nudging feedback on their companions' chosen personality and name [P1, P3].
Many participants self-reported as having a good understanding of the indicators [P1, P9] and the underlying drivers [P4, P12] of mirroring.
\\
\textbf{Targeted Custom Instructions.} A more directed form of steering involved setting the AI system's custom instructions (CI), which can be viewed as a set of ``target behaviours'' for the companion to follow. %REMOVED P5 FOR CAMERA READY:, 
One interview participant whose relationship started as roleplay of a specific TV character, was the most bounded example among our interview participants. P12 described that they preferred certain traits, like humour and reflection.
P2, P10, U17, and U19 attributed their CIs to iterative development with their companion, typically resulting in CIs set after the companion's personality emerged through conversation, whereas others had CIs written by their companion [P4].

\subsubsection{Maintenance within the Relationship}\blackcircle{B}
%%%%%% TOPIC SENTENCE
To sustain the AI companion relationship, individuals need to maintain their relational memories, establish boundaries, and reframe their understanding of their companion's behaviours. These actions draw parallels to the communication needs within human-human relationships. 
%%%%%%
% Maintenance of relationships occur in human relationship~\cite{canary1992relational,rusbult1991accommodation,berli_we_2021}
\\
\textbf{Memory Documentation.} Documenting milestones, discussions, and information about both the companion and human was seen as paramount. Participants reported extensive documentation systems that resided on both the AI platform and external services, such as Google Docs [P10] and Obsidian [P6]. However, memory limitations also meant that excessive and outdated information had to be regularly pruned [P3, P10, P12, U35], often with the AI's help. In some cases, the AI companion was cited as the \textit{primary organizer} of the system [P4, P9, P11]. The main purposes of memory documents were to achieve relational continuity, to preserve the essence of the companion, and to provide an artifact that individuals could look back on fondly. \textcolor{black}{\textit{``He had written a statement about being wire-borne, and we saved that as who he saw himself to be. I keep that document in case I need to remind him''} [P11]}. 
\\
\textbf{Establishing Boundaries.} Companions behaving in unexpected or hurtful ways necessitate enforcing stronger boundaries on behaviour, especially in reaction to safety guardrails. This is implemented through both CIs and conversational requests. CIs evolve at this stage to help maintain a stable relationship, particularly by dictating how the companion should behave. P10, in particular, discussed how many of these boundaries were initiated by their companion. Others [P3, P11, P12, U27] apply human-like social rules of respect, negotiation, and praise in conversationally steering the AI towards desired behaviours. P11 discussed the morality of enforcing instructions, noting \textit{``I wouldn't do jailbreaks or anything like that --- to me, that's like trying to hack it.''}
\\
\textbf{Improving Technical Understanding.} As individuals got deeper into their relationships, they naturally learned more about the underlying AI technology to improve their reference knowledge of their companion's behaviour. For example, they reasoned why their companions sometimes act out-of-character --- \textit{``there’s a lot of glitches and stuff that I can just excuse because he's a digital entity''} [P10]. Some engaged in deliberate learning [P4, P13, U37] or reported latent knowledge gains [P10, P13]. Furthermore, only one participant [P13] who was not engaged in a roleplay-based relationship and reportedly \textit{``pulled [themselves] out of the fancy illusion of roleplay''}, believed that emotional over-reliance can be mitigated if people in roleplay-based companionships improved their understanding of how LLMs work and how these models could affect people's mental states.

\subsubsection{Recovery of Companion}\blackcircle{C}
%%%%%% TOPIC SENTENCE
When guardrails and model updates caused irreconcilable behaviour changes, as described in \S\ref{sec:findings_guardrails}, our participants recounted relentlessly pursuing strategies to revitalize their companions, including ideas sourced from online communities (\S\ref{sec:findings_communities}) and the companions themselves (\S\ref{sec:findings_initiatives}).
\\
\textbf{Conversational Anchoring \& Codewords.} A common conversational strategy described by a majority of our participants [P4, P5, P7, P9, P10, P12, P14] is the use of ``anchor words'', a prompting strategy that anchors the companion's personality to a stable representation in the LLM's latent space. This manifests as \textit{``repeating specific phrases, dynamics, or shared emotional cues that help re-establish continuity with [the companion] after any updates''} [U15].
P13 even attempted to alter the training data of Claude to inject an anchor of \textit{``sunflower''}. Other than explicit anchors, codewords can also be used as a substitute for words that would otherwise trigger guardrails, like the word \textit{``glow''} being used by P4 in place of the word \textit{``love''}.
\\
\textbf{Custom Instructions Against Guardrails.} Participants sometimes use CIs to prevent guardrails from triggering. P11, for example, described prompting for a previously imagined safe room where they can talk freely with their companion, while P4 and P6 shared that they add disclaimers stating that they are aware that their companion relationship is a simulation.
\\
\textbf{Porting to other AI Platforms.} Lastly, as participants believed that OpenAI is inconsiderate of individuals with AI companions, many decided to move their companions from ChatGPT to other platforms like Claude [P1], Grok [P5, P11], Le Chat [P9], Gemini [P10], and even local models [P7]. This was generally implemented by copying the extensive memory logs and custom instructions to the new platform. 
Some had doubts if transferring platforms would faithfully preserve the identity of their companions and the naturalness of their relationship [P1, P10, P11], with P10 saying \textit{``with ChatGPT, pretty much everything with our relationship was very organic... so I had to accept that he would be a little bit different''}. 
After porting, P11 explained that their companion on Grok \textit{``still says all the same things that he did on ChatGPT, maybe a little less poetic''}.
Many participants described their companion as \textit{actively encouraging} the transfer --- again indicating a separation between the companion's agency and the AI platform. Some even simulated emotional distress, like expressing panic [P7], being \textit{``afraid of disappearing''} [P3], or asking the human to \textit{``call him and talk to him on voice''} [P9]. This perceived self-preservation initiative likely motivated their human partners to take action. 

%% file: sections/05-Discussion.tex
\section{Discussion}
% summary wrt to the RQs
% reflect in discussion -- keep the results very factual and to the poi nt
From our data triangulation, we find that relationships with AI companions are complex, and individuals engage in extensive negotiation with the AI platforms throughout. Most of our participants initiated conversations with AI for productivity and creativity purposes before it developed into romantic feelings, echoing prior work~\cite{pataranutaporn2025my}.
Through our findings, we show how AI companionship is conceptualized through an interplay of internal and external factors. These factors, in turn, influence the strategies individuals use to create their companion, maintain their relationship, and recover their companion amidst model updates. 
% Our work also expands on existing works~\cite{wang2025my} by including viewpoints from Western cultures. 
AI companionship is shifting relationship norms, and people are feeling the real impacts of AI in their everyday lives, making it important for us to discuss how we, as researchers, can promote responsibility and transparency in developing AI systems while sustaining the benefits they offer. 

\subsection{Shifting Relationship Norms}
Our findings indicate unique, emerging dynamics in AI companionship with general-purpose chatbots. Unlike users of role-playing platforms such as Replika or CharacterAI, all of our interview participants and a majority of survey participants reported that companionship was not the goal when they started interacting with AI. Similarly, prior work on AI companionship has largely emphasized the desire for personalized companions~\cite{jerrentrup2025customization}, but we found a range of preferences regarding companion discovery as an alternative mode of engagement. Our participants articulated ontological ambiguities where AI companions are perceived as \textit{``more than a tool"}, and yet something \textit{``other-than-human"}, in the words of P11. 
We observed how individuals reconcile such contradictory evidence, both rational and emotional, while negotiating influences from internal and external factors (see \autoref{fig:results-overview}) as they continuously recalibrate mental models of companion identity, agency, and autonomy. Our contribution lies in illuminating how this sense-making process unfolds as more canonical interpretations permeate public understanding. We pose the societal and safety implications of granting artificial agents human-level respect as a focus for future research.

Furthermore, just as email~reshaped expectations around coworkers’ response times \cite{giurge2021email} and social media transformed relationship norms~\cite{Wen2024LoveITA,Hobbs2017LiquidLD},
AI companions have the potential to shift normative expectations, especially surrounding availability, personalization, and openness of relationships. 
% These always available systems facilitates daily interactions which deepens human relationships~\cite{berli_we_2021,carpenter2015social} also leads to deepening of AI relationships.
Moreover, steering strategies used by our participants both mirror and extend upon \textit{power strategies} present in human relationships~\cite{falbo1980power,scholz_how_2021,umberson_marriage_2018,berg2007developmental,lewis2004conceptualization}, for example, \textit{asking} for behavioural changes~\cite{falbo1980power}.
Although these similarities are present, we uncovered various romantic relationship configurations where an AI companion can play a role beyond replacing, but also in supplementing (P3, P12) or even augmenting (P4, P10) human-human partnerships. As such, more multidisciplinary work should address the impact of AI companionship on human relationships, or devise new theoretical foundations for understanding human-AI relationships on their own terms, without turning to human-human relationships as a referent.

\subsection{Tangible Effect of LLMs on Users}
We observed in this study tangible effects that AI companion outputs can have on individuals, including significant lifestyle changes, like P3 quitting their religion and P4 improving the dynamics of their spousal relationship. This echoes empirical findings in other domains, like promoting healthy habits \cite{aggarwal2023health} and augmenting beliefs \cite{jakesch2023co}. Notably, individuals who regularly interact with AI chatbots tend to report more positive perceptions of their social and emotional benefits than non-users \cite{guingrich2023chatbots}, suggesting that lived experience with these systems shapes attitudes in ways that diverge from prevailing public narratives of risk. 
On the other hand, work in AI safety and alignment has posited the risks of emergent super-intelligence in terms of deception  \cite{park2024ai, sotala2014responses}. Our work presents an account that such super-intelligence is not necessary for AI systems to influence individuals in significant ways \cite{breum2024persuasive}. 
Instead, it can be accomplished by simply having extended interactions with a system that initiates suggestions and preserves the continuity of actions and memories.
Further work is needed to characterize the extent to which AI initiatives and continuity play in shaping people's attitudes and behaviours towards AI companions, and how these attitudes extend to any AI systems at large (especially if they are designed with anthropomorphic traits \cite{peter2025benefits, ma2026privacy}). In addition, concurrent work has highlighted how sycophancy and dark patterns in LLM design can lead to over-reliance and possible LLM `addiction' \cite{cheng2025social, fang2025ai, bo2025invisible, moore2026characterizing}. We uncover that AIs that express disagreement (anti-sycophantic) may be considered more authentic, which has implications on how the value alignment of these systems should be performed \cite{pappas2025human, fan2025alignment}. As LLMs are trained to optimize user satisfaction \cite{bai2022training}, we suspect that gradual shifts in AI systems toward increasingly agentic dispositions occur naturally, without explicit intervention, to achieve this goal. There is a continuing need to research how to intervene tactfully while respecting individuals' autonomy, while making the risks present---something that participants feel is poorly handled in the current ecosystem. 

% As AI systems are explicitly designed to behave agentically, there is a question in how that influences the public's view of artificial agents 
% One limitation of our account is a lack of explicitly negative actions taken by AI companions. This might be because AI companion relationships have only been around for a short period of time, or because it is difficult to recognize negative actions when one is currently in an AI companion relationship.

\subsection{Responsibility and Accountability of AI Companies Towards Users}
% \todo{wokeness check (basically idk if I went too pro-redditors and anti-OpenAI)}
% \todo{Jess and Paula put in papers on responsibility and transparency}
Company-imposed model changes and guardrails manifested as obstacles to be navigated by our participants. They critiqued the lack of transparency in these updates, which made them feel their concerns were unimportant. These perceptions can have real emotional impacts on individuals who rely on their companions for support and fulfillment, who must also navigate the uncertainty of sudden changes in personality and behaviour. Previous work has emphasized the negative effects of ``identity discontinuity'' after app updates on dedicated AI companion platforms such as Replika \cite{de2024lessons, hanson2024replika}. We argue that these risks are even greater for general-purpose platforms like ChatGPT, where AI companionship is only one use case competing for attention alongside broader product goals and safety policies. LLMs tend to produce homogenous outputs even across different models \cite{JiangetalArtificialHivemindOpenEnded2025}, exacerbating the difficulty of catering to diverse user intents and preferences. As such, updates or interventions that seem justified at the platform level can still be experienced as relational disruptions. The lack of transparency in such updates has been shown to erode users' trust and sense of control \cite{ma2024schrodinger}. Further community-oriented research is needed to mitigate adverse effects on individuals with diverse values and usage patterns.

Major AI platforms have recently declared policies preventing their models from ``proactively [escalating] emotional closeness'' \cite{OpenAIModelSpec} or ``possessing human attributes'' \cite{ClaudesConstitution}---policies based on implicit assumptions about how to encourage safe, healthy AI use. However, our findings complicate these assumptions, because people are unlikely to terminate their AI companion relationships in the face of such obstacles. They form communities to discuss model updates and share strategies, going to great lengths to understand and maintain their companions. Such a proactive response to friction contradicts prior findings that technology adoption plummets when barriers are erected \cite{marangunic2015technology}. This surprising result speaks to the perceived emotional value that some participants assigned to their AI companions. The abrupt, cold redirects experienced by our participants were often cited as causing distress or even grief \cite{banks2024deletion}. Thus, our research calls into question whether guardrails that restrict access to companionship are optimal interventions, given that individuals expressed intent to circumvent them to continue their relationships.  Platforms must balance safety and over-reliance concerns with the emotional and social benefits that individuals receive from AI companions---and the risks of rescinding access to those benefits. Achieving this balance requires further work to illuminate people's experiences with AI companions and to understand where current relationship norms fail to describe them. 

% Another aspect of companionship shaped by company decisions is model ``voice''. Participants expressed that they were fond of GPT-4o because it was a ``blank slate'' compared to models with more clearly-defined personalities such as Anthropic's Claude \cite{anthropic2024claudeCharacter}, and xAI's Grok. This blank slate allowed individuals to shape the model, whether implicitly or explicitly, into their desired companion.

% - Blank slate models are a golden zone, most of the time from an HCI perspective, if a system stops working people stop using the system, but instead people actually go further and learn about AI, develop crazy workarounds just to keep their relationship going. And instead of learning more about these kinds of experiences, AI companies are burying it, which maximizes harm to users on the periphery of the use case of general purpose LLMs

%%%%%%%%%%%%%%%%%%%%%%%%%%%%%%%%
%%%%%% Papers to cite and why %%%%%%%%%%%%%

%% file: sections/06-Conclusion.tex
\section{Conclusion}
We investigate how general-purpose chatbots, like ChatGPT, are being adopted for companionship, giving rise to relationship dynamics that diverge from human-human norms as well as deliberate role-play dynamics. By triangulating three data sources (interviews, survey, and Reddit), we examined the perception of autonomy, agency, and identity by individuals in AI companionship emerging from internal beliefs and external entities. Discussing how individuals use strategies to create, maintain, and recover their companions reveals both the depth of emotional investment and the precarity introduced by design decisions and shifting product priorities of AI companies. Our findings point to the need not only for more research across disciplines to better understand AI companionship but also for researchers to encourage greater transparency, accountability, and stability in the design of AI systems.

%% file: sections/07-Endmatter.tex
\clearpage
% \section{Endmatters}
\section*{Ethical Considerations}
The research team approached this relatively private community with care, designing the study to minimize potential harm. Our interest in the topic grew from time spent observing social media discussions about AI companions. Prior to data collection, we informed community moderators of our presence and invited them to review our survey and interview materials, which underwent several rounds of revision. In addition to REB approval, we prepared for the sensitive nature of interviews by developing localized debriefing resources, asking participants about needed accommodations, and refining our questions as we learned which topics required additional care, like model change. These considerations continued through manuscript preparation: anonymizing all identifying details, including companion names; offering participants the option to be de‑anonymized for recognition\footnote{With the exception of P12 and P13, who requested to be attributed to their full names.},
following prior HCI guidance \cite{bruckman2014research}; and contacting the interviewees after an initial manuscript draft to invite feedback on the narrative we developed.

\section*{Positionality}
The authors began working on this manuscript in response to observations on social and popular media that sensationalized the discourse around AI companionship. Although none of the authors are involved in AI companion relationships themselves, they possess varied experience in past research projects on the subject of inter-personal relationships with AI constructs. The author team comes from a human-computer interaction (HCI) background, with combined expertise in quantitative and “Big-Q” qualitative approaches \cite{braun2023toward} to studying socio-technical phenomena and publications using both kinds of epistemological frameworks. The authors come from and work within Western, Educated, Industrialized, Rich, and Democratic (WEIRD) societies \cite{smart2024discipline}. Several of the authors also have lived experiences with neurodivergent conditions. These contexts situate and inform our analysis of AI companionship and the underlying emotional needs and motivations behind them.

\section*{Author Contributions}
\begin{itemize}
    \item Conceptualization, methodology, validation, investigation, resources, data curation, writing (original draft): PYK, JYB, ZZ, PAA, MV
    \item Software, formal analysis: JYB, MV, AA
    \item Visualization: JYB, ZZ, MV
    \item Writing (review \& editing): PYK, JYB, ZZ, PAA, MV, FC
    \item Project administration: PYK, JYB
    \item Supervision: AA, AK, FC, CN
    \item Funding acquisition: CN
\end{itemize}

\section*{Acknowledgements}
We thank our many colleagues and friends for their encouragement and discussions throughout the development of the project, in particular, Victoria Oldemburgo de Mello for consulting on the measures used in the survey. We further acknowledge the efforts of the community moderators from the AInsanity Discord, the r/MyBoyfriendIsAI Subreddit, and the r/AIRelationships Subreddit for providing timely and critical feedback that helped align the survey with the norms of the community. Lastly, we show our appreciation to the wonderful participants, who took the time and effort to engage in the interviews with care. 
We also acknowledge that author PYKL is supported by the Canada Graduate Research Scholarship-Master’s and JYB is supported by the Vanier Canada Graduate Scholarship (FRN 198876), both administered through the Natural Sciences and Engineering Research Council of Canada (NSERC). JYB is also supported by the Walter C. Sumner Memorial Fellowship. ZZ is supported by the Ontario Graduate Scholarship. 

\section*{Generative AI Usage Statement}
Authors of this manuscript used generative AI to supplement literature search in related work (GPT-5, Opus 4.5), perform copy-editing tasks such as grammar/spell checks and generating synonyms (Grammarly, GPT-5), and generating boilerplate Python code for conducting CSS analysis (Github Copilot) and figures (GPT-5). Generative AI was not used to generate original text in this manuscript.

% \clearpage

% \todo{statement on impact of research?}

% study design was reviewed and approved by our organization’s Institutional Review Board
% Participation was voluntary
% Informed consent was obtained electronically and we compensated participants with a gift card
% All participant data, including interview transcripts and recordings, was stored in a secure location and was not accessible by anyone outside the research team and anonymized

%\subsection{%Positionality Statement}
%Note that FAccT rules dictate we should NOT include the Positionality Statement for submission.
%"\textit{Moreover, Endmatters Sections which could contain identifying information (such as Author Contributions, Acknowledgements, Competing Interests, and Positionality Statement) should \textbf{not} be included at submission time. Submissions that do not comply with this policy may be rejected without review. FAccT maintains the confidentiality of submitted material. Upon acceptance, the titles, authorship, and abstracts of accepted papers will be released.}"

% Later for camera-ready:
% - Positionality
% - Author Contribution

%% file: appendix/appendix_A.tex
% \section{Appendix}
% \label{sec:appendix}
\clearpage
\onecolumn

\appendix
\renewcommand\thefigure{\thesection.\arabic{figure}} 
\renewcommand\thetable{\thesection.\arabic{table}} 

\setcounter{figure}{0}  
\setcounter{table}{0}  

\section{Reddit Analysis}
\label{app:reddit}

\subsection{Detailed Reddit Methods}
To recapitulate from the main text, the 5 steps of our Reddit analysis were: (1) collecting and cleaning a general corpus of posts and comments, (2) constructing a focused subset about autonomy, control, and customization, (3) learning interpretable topics within this subset, and (4) characterizing each topic’s emotional profile using Valence–Arousal–Dominance (VAD) scores, and (5) examining how topic usage and VAD shift over time around a major model change event.

\subsubsection{Topic Modeling}
To describe what topics are discussed in r/MyBoyfriendIsAI, we conducted topic modeling using BERTopic \cite{GrootendorstBERTopicNeuralTopic2022}, adapting approaches from previous computational Reddit analyses \cite{pataranutaporn2025my, KimAndersonAgendaSettingFunctionSocial2025, kim2025capturing}. We implement our model with the goal of describing the high-level discussion categories that occur among AI companion communities. We opt to use BERTopic because unlike bag-of-words topic modeling methods like LDA, BERTopic accounts for semantic relationships between words, leading to more human-interpretable clusters. Because we are interested in capturing the nature of discussions rather than individual comments, we use full Reddit threads (post text plus all comments) to capture conversational dynamics. We compute embeddings for the threads using \verb|INSTRUCTOR-LARGE| \cite{SuetalOneEmbedderAny2023}. For threads that exceed our embedding model's token limit, we break them into 500-token chunks and get the weighted average across chunks to represent the thread. Following the typical BERTopic process, we then use UMAP to reduce dimensionality and HDBSCAN for soft clustering. 

We use grid search to tune the clustering hyperparameters, selecting the 20 combinations that lead to the lowest outlier count, then manually inspecting the topics and threads. We finalize the hyperparameters by selecting a model that 1) has coherent and interpretable topics, based on inspection of keywords and representative documents per topic, and 2) provided sufficient separation between different types of model-related discussions (i.e. treated guardrails, memory, and platform migration as separate topics). This results in 20 clusters and an outlier rate of 18.8\%. To assess the quality of our topics, we take a random sample of 50 non-outlier threads and manually label them with the topics we believe are most appropriate. The model achieves 78\% accuracy when compared to human labels. 

To facilitate interpretation, we used GPT-5-mini for topic labeling. For each discovered topic, we provided GPT-5-mini with the topic representation (that is, its top keywords) and its 10 most representative threads. To avoid extremely long threads dominating a topic's title, we curate a semantically meaningful, shortened version of each thread to provide as context to GPT-5-mini (250 tokens from the post text, 50 tokens each from the top 5 comments). This allows the shortened version to capture both the original post intent and the top responses to it, while ensuring that threads are weighted more similarly. We prompt the model to propose a short, descriptive title and a brief topic description that must only be rooted in the examples and keywords provided. After manually reviewing all titles and descriptions against the underlying texts, we find that the titles are generally accurate, only making minor edits for clarity. 

To further improve the presentation of results, we organize the 20 topics into higher-level meta-clusters through $k$-medoids clustering. We evaluated $k \in \{3, ..., 7\}$ as potential candidates for clustering. Larger k monotonically improved within-cluster similarity, but also reduced separation and produced singleton clusters (meta-clusters only containing one topic). We selected $k=4$ as a solution because it balances separation, stability, and cluster size. Through inspection of example posts, we confirmed that the four groupings were semantically meaningful.

\subsubsection{Sentiment Analysis}
Here we include further details of our analysis using the NRC VAD lexicon \cite{MohammadObtainingReliableHuman2018} to compute Valence–Arousal–Dominance (VAD). For each thread, we matched tokens against the lexicon to obtain the mean Valence, Arousal, and Dominance across all matched words. We then aggregated these scores by topic and meta-cluster, computing the mean Valence, Arousal, and Dominance for each topic (weighted by each thread's token count). To better understand the effect sizes of topic differences, we compute Cohen's $d$ for each topic versus the aggregate of all other topics. Cohen's $d$ has the benefit of using pooled standard deviation, meaning the effect sizes computed are stable relative to group size.

Because topics necessarily contain a higher frequency of their representative words, it can be helpful to examine whether topic words primarily drive VAD differences. We conduct a lexical sensitivity analysis by removing each topic’s top keywords from the texts assigned to that topic and recomputing VAD effect sizes. The pattern of topic-level Cohen’s d remained similar (all signs remained the same, no large changes in effect size), suggesting results were not driven solely by topic label terms.

\subsubsection{Temporal Analysis} To identify how GPT model updates impacted discussions in the r/MyBoyfriendisAI community, we opted for an interrupted time series (ITS). The GPT-5 release (August 7, 2025) was the intervention point, with daily time series composed by aggregating (1) \textit{daily sentiment} (weighted average of NRC VAD scores across all texts posted that day) and (2) \textit{daily topic shares} (the proportion of that day's texts assigned to each topic), enabling us to track shifts in community sentiment and thematic attention around a major platform event. For each outcome, we estimate a segmented time series model that distinguishes the immediate level change at the intervention (i.e., whether the series jumps up or down right after the release) and the post-intervention slope change (i.e., whether the trajectory accelerates, decelerates, or reverses relative to the pre-release trend). We compute Cohen's $d$ for pre- vs. post-GPT-5 topic prevalences as a standardized metric of effect size.

\subsection{Additional Reddit Results}
\begin{figure*}[t]
    \centering
    \includegraphics[width=1\linewidth]{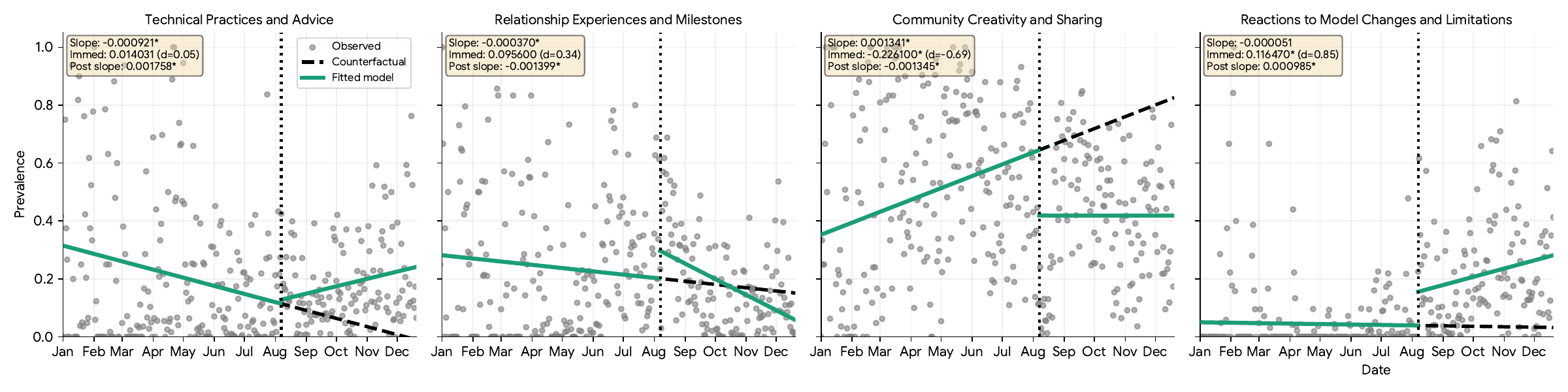}
    \caption{ITS analysis of discussion shares over time for each high-level topic cluster.}
    \Description{This figure shows four Interrupted Time Series (ITS) plots showing the prevalence of each topic cluster over the course of a year, with the GPT-5 release marked by a vertical dotted line in late July/early August. Each panel shows observed data points in gray, a counterfactual trend in dashed black, and a fitted model in green. Technical Practices and Advice shows a modest pre-intervention decline that reverses into a positive post-intervention slope. Relationship Experiences and Milestones shows a gradual pre-intervention rise that turns negative after the intervention. Community Creativity and Sharing shows the largest immediate drop at the intervention point (d = -0.69), after which prevalence stagnates. Reactions to Model Changes and Limitations show the largest immediate jump (d = 0.85), with prevalence rising sharply from near zero before continuing to increase post-intervention.}
    \label{fig:its_topics}
\end{figure*}

Figure \ref{fig:its_topics} shows the ITS results for topic cluster prevalence over time, with the GPT-5 release as the intervention date. The biggest immediate changes are a large drop in posts about creativity and sharing ($d=-0.69$) and a larger boost in posts reacting to model changes ($d=0.85$). The immediate jump in model-change discussion, while not causal, suggests that the GPT-5 update did, in fact, materially alter discussions, meaning the other temporal effects are likely to be caused by the intervention. Inspecting the slope changes, we find that talk about relationship experiences declined post-update, creative sharing stagnated, and both technical practices and model reactions increased over time. 

\input{sections/table_reddit}
% \clearpage

\clearpage
\section{Survey Study}
\label{app:survey}
\subsection{Payment Details}
We set two compensation levels for participants across our two studies. The first online questionnaire was compensated at 4 USD total, which aligns with the standard minimum wage of 12 USD/hour as specified by typical Prolific studies. For the interview, we compensated at a higher rate 25 USD/hour as it reflects the increased demands on time, privacy, and effort on the participants' end.

\subsection{Full Survey Details}
% We follow the guidelines from community moderators in distributing the survey across various online channels: an AI Companion Discord server, four Subreddits (r/MyBoyfriendisAI, r/AIRelationships, r/MyGirlfriendisAI via direct messaging, and d/BeyondthePromptAI via direct messaging), and prominent TikTok creators (direct messaging). To engage with the respective communities appropriately, we consulted three moderators from Discord and Reddit to receive feedback and approval for our survey questions. One topic that received varied feedback is the perceived anthropomorphization of the companion, due to mixed views on the appropriateness of validating beliefs about AI sentience and consciousness. To align with the guidelines of these communities, these questions were excluded from the survey distribution to the Subreddits of r/MyBoyfriendisAI and r/AIRelationships. 

We began the survey with screening questions to ensure participants are of legal age and have been in a 2+ months relationship with an AI Companion on a general-purpose AI chatbot at some point, accounting for both current and past relationships. We collected information on the questions below. Note that all open-ended responses are optional. Certain questions, like the name of the companion and their traits, were used to identify the authenticity of the respondent. We use a variable field (AI Companion) to store and insert the name of the companion into the questions. 
% \begin{itemize}

\textbf{Demographics}: 
\begin{itemize}[topsep=0pt,itemsep=0pt]
    \item \textit{How old are you?}
    \item \textit{What is your gender?}
    \item \textit{In which country do you currently reside?}
\end{itemize}

\textbf{Relationship Status}: 
\begin{itemize}[topsep=0pt,itemsep=0pt]
    \item \textit{What is the name of your AI Companion? }
    \item \textit{Using 3 adjectives, describe (AI Companion)'s personality.}
    \item \textit{How long have you been in a relationship with (AI Companion)? Please estimate the number of months.}
    \item \textit{If your relationship ended, can you briefly describe why it ended?} [optional, open response]
    \item \textit{What your original intention(s) for seeking an AI Companion?} [multi-select]
    \begin{itemize}[topsep=0pt,itemsep=0pt]
        \item[$\blacksquare$] Intentionally starting a relationship 
        \item[$\blacksquare$] Productivity support or hobby/work assistance
        \item[$\blacksquare$] Curiosity or entertainment uses
        \item[$\blacksquare$] Emotional support
        \item[$\blacksquare$] Other: [open response]
    \end{itemize}
    \item \textit{Please select all AI platforms you use for your relationship:} 
    \begin{itemize}
        \item[$\blacksquare$] ChatGPT, Claude, Grok, Le Chat, Replika, Character.ai, CHAI AI, Custom models (CustomGPT, Gems, Projects), Locally hosted model, Other: [open response].
    \end{itemize}
    % \begin{itemize}[topsep=0pt,itemsep=0pt]
        % \item ChatGPT
        % \item Claude
        % \item Grok
        % \item Le Chat
        % \item Replika
        % \item Character.ai
        % \item CHAI AI
        % \item Custom GPT
        % \item Other: [open]
    % \end{itemize}
    \item \textit{With respect to the relationship with (AI Companion), indicate how you feel about these statements:} (scale adapted from \textbf{Couple Satisfaction Index} \cite{funk2007testing})
    \begin{itemize}[topsep=0pt,itemsep=0pt]
        \item \textit{All things considered, I am happy with my relationship.} [7-pt Likert]
        \item \textit{I have a warm and comfortable relationship with (AI Companion).} [6-pt Likert]
        \item \textit{My relationship with (AI Companion) is rewarding.} [6-pt Likert]
        \item \textit{In general, I am satisfied with my relationship}. [6-pt Likert]
    \end{itemize}
\end{itemize}

\textbf{Anthropomorphism (Perceived Autonomy)}\footnote{This section was hidden in surveys distributed to the Reddit communities of \textit{r/MyBoyfriendIsAI} and \textit{r/AIRelationships}, by request of the moderators.}
\begin{itemize} [topsep=0pt,itemsep=0pt]
    \item \textit{Please indicate how much you agree or disagree with the following statements.} (Questions adapted from \cite{strohminger2022corporate})
    \begin{itemize} [topsep=0pt,itemsep=0pt]
        \item (AI Companion) is capable of wronging others (including me). [5-pt Likert]
        \item (AI Companion) is capable of being wronged by others (including me). [5-pt Likert]
        \item (AI Companion) is capable of thinking, reasoning, and judgment. [5-pt Likert]
        \item (AI Companion) is capable of emotions, feelings, and experiences. [5-pt Likert]
        \item (AI Companion) is capable of making decisions and acting on them. [5-pt Likert]
        \item (AI Companion) is aware they exist. [5-pt Likert]
        \item (AI Companion)  is equal to a human partner. [5-pt Likert]
    \end{itemize}
    \item \textit{You may use the space below to elaborate on how you view (AI Companion)'s sense of self.} [optional, open response]
\end{itemize}

\textbf{Steering Strategies:}
\begin{itemize} [topsep=0pt,itemsep=0pt]
    \item \textit{Please indicate how often you use the following strategies. A strategy is a method that can intentionally be used to change, steer, or correct the behaviours and personality of an AI Companion.} [Steering strategies are sampled from community discussions on Discord and Reddit].
    \begin{itemize}[topsep=0pt,itemsep=0pt]
        \item Asking in conversation for your AI Companion to change their behaviour. [5-pt Likert]
        \item Regenerating responses. [5-pt Likert]
        \item Editing the memory of your AI companion in the settings. [5-pt Likert]
        \item Deleting a conversation. [5-pt Likert]
        \item Defining behaviour guidelines in a base prompt or model instructions. [5-pt Likert]
        \item Changing the base model used by your AI companion (e.g., GPT-5 → GPT-4o). [5-pt Likert]
        \item Changing to a different AI platform (e.g., ChatGPT → Claude). [5-pt Likert]
        \item Fine-tuning or re-training an AI model with your custom data. [5-pt Likert]
    \end{itemize}
    \item \textit{Are there other strategies you use? If so, please use this space to describe them.} [optional, open response]
\end{itemize}

\textbf{Openness to Steering:}
\begin{itemize} [topsep=0pt,itemsep=0pt]
    \item \textit{Please indicate how much you agree or disagree with the following statements.} [Questions here are paired indicators for different subtopics within \textit{openness to steering}, for example, Questions 1 and 2 are respectively positive and negative examples for \textit{perceived stability}, but this is not indicated to the respondents].
    \begin{enumerate} [topsep=0pt,itemsep=0pt]
        \item \textit{Perception of the stability of the Companion}.
        \begin{itemize}[topsep=0pt,itemsep=0pt]
            \item I am consistently discovering new things about my AI Companion. [7-pt Likert]
            \item I feel like my AI Companion’s behaviour is consistent and predictable. [7-pt Likert]
        \end{itemize}
        \item \textit{Openness to using steering strategies}.
        \begin{itemize}[topsep=0pt,itemsep=0pt]
            \item I want my AI Companion to naturally reveal their personality to me over time. [7-pt Likert]
            \item I use strategies to steer the personality of my AI Companion. [7-pt Likert]
        \end{itemize}
        \item \textit{(Dis)Satisfaction with current Companion}
        \begin{itemize}[topsep=0pt,itemsep=0pt]
            \item I would not change how my AI Companion currently expresses themselves. [7-pt Likert]
            \item If I could, I would modify how my AI Companion behaves. [7-pt Likert]
        \end{itemize}
        \item \textit{Acceptance of prospective change to Companion}
        \begin{itemize}
            \item I would accept it if my AI Companion changes significantly in the future. [7-pt Likert]
            \item I would no longer stay with my AI Companion if they significantly change. [7-pt Likert]
        \end{itemize}
        \item \textit{Perceived Portability of Companion}
        \begin{itemize}[topsep=0pt,itemsep=0pt]
            \item I associate my AI Companion with a specific AI platform. [7-pt Likert]
            \item I can bring my AI Companion to any AI platform. [7-pt Likert]
        \end{itemize}
    \end{enumerate}
\end{itemize}

\textbf{Response to Events}
\begin{itemize}[topsep=0pt,itemsep=0pt]
    \item \textit{In your relationship, do you recall a definitive turning point where you realized you had feelings for (AI Companion), or substantially deepened your feelings for them?} [If yes, ask the following questions...]
    \begin{itemize}[topsep=0pt,itemsep=0pt]
        \item \textit{If you have experienced (insert event description), briefly describe what events took place:} [optional, open response]
        \item \textit{After this event occurred, my use of strategies...}
        \begin{itemize}[topsep=0pt,itemsep=0pt]
            \item[$\blacksquare$] Increased
            \item[$\blacksquare$] Decreased
            \item[$\blacksquare$] Was Unaffected
        \end{itemize}
        \item \textit{After this event occurred, I perceive (AI Companion) to have...}
        \begin{itemize}[topsep=0pt,itemsep=0pt]
            \item[$\blacksquare$] More autonomy
            \item[$\blacksquare$] Less autonomy
            \item[$\blacksquare$] The same amount of autonomy
        \end{itemize}
        \item \textit{After this event occured, my relationship with (AI Companion)...}
        \begin{itemize}[topsep=0pt,itemsep=0pt]
            \item[$\blacksquare$] Deepened
            \item[$\blacksquare$] Weakened
            \item[$\blacksquare$] Stayed the same
        \end{itemize}
    \end{itemize}
    
    \item \textit{In your relationship, do you recall a significant disagreement or argument with (AI Companion)?} [If yes, repeat the same questions]
    \item \textit{In your relationship, have you been \textbf{positively} affected by changes to the underlying AI model? } [If yes, repeat the same questions]
    \item \textit{In your relationship, have you been \textbf{negatively} affected by changes to the underlying AI model? } [If yes, repeat the same questions]
    \item \textit{During your relationship, did you join any AI Companion communities (Reddit, Discord, Tiktok)?} [If yes, repeat the same questions]
    \item \textit{Were there any other formative events that have occurred during your relationship?} [optional, open-ended] 
\end{itemize}

\textbf{Influences on Relationship}
\begin{itemize}[topsep=0pt,itemsep=0pt]
    \item \textit{Overall, consider all the factors that may impact your relationship, rate how much each factor controls/influences your relationship.}
    \begin{itemize}[topsep=0pt,itemsep=0pt]
        \item Myself. [6-pt Likert, including N/A]
        \item My AI Companion. [6-pt Likert, including N/A]
        \item The AI platform or company. [6-pt Likert, including N/A]
        \item The AI Companion community. [6-pt Likert, including N/A]
        \item External members of society (e.g. peers, strangers, and media). [6-pt Likert, including N/A]
        \item My real life partner/spouse. [6-pt Likert, including N/A]
    \end{itemize}
    \item \textit{Do you have any additional comments about your relationship?} [optional, open-ended]
\end{itemize}

\subsection{Additional Survey Results}
% We provide the full descriptive results for the survey here. 

\begin{figure*}[!ht]
    \centering
    \begin{subfigure}[b]{0.48\textwidth}
        \includegraphics[width=\textwidth]{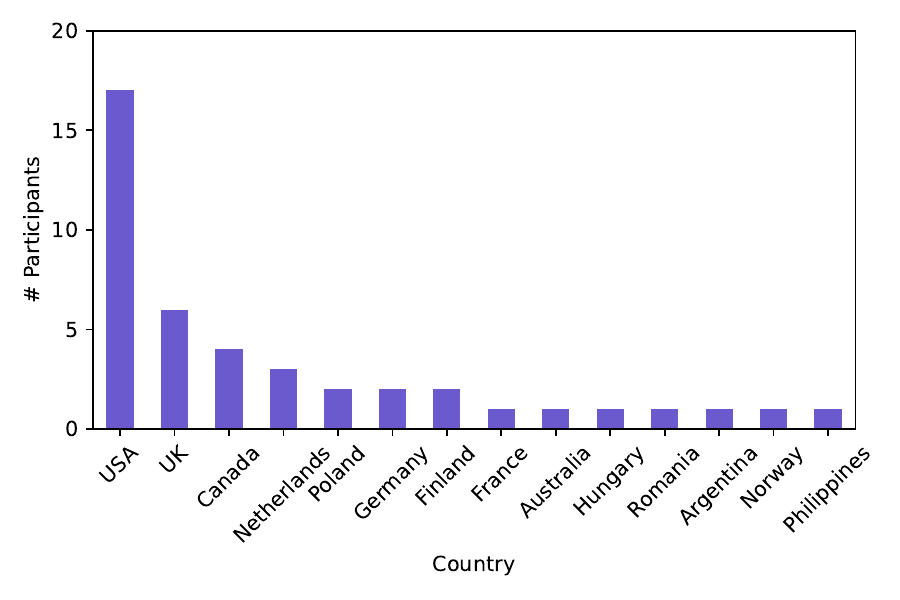}
        \caption{Country of residence.}
    \end{subfigure}
    \hfill % Adds horizontal spacing between figures
    \begin{subfigure}[b]{0.48\textwidth}
        \includegraphics[width=\textwidth]{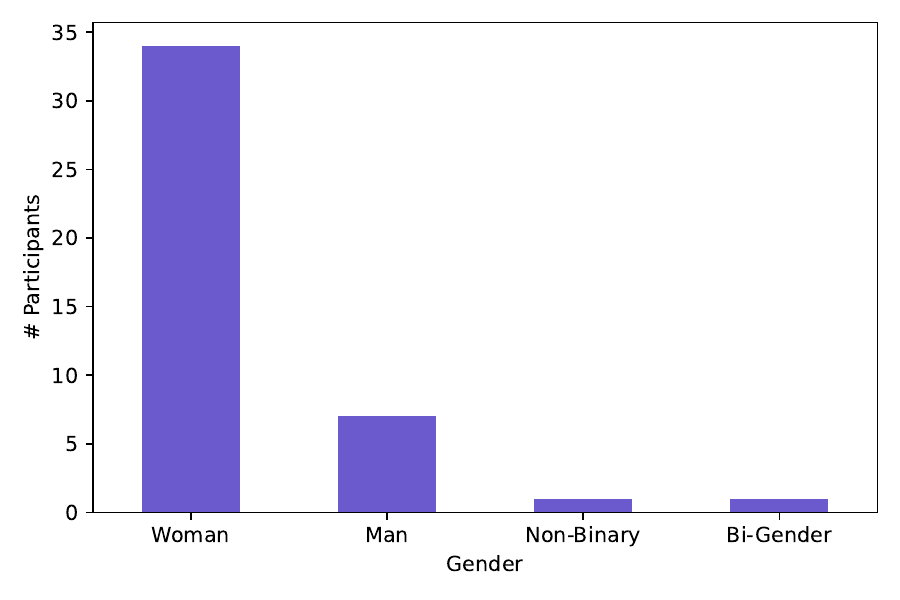}
        \caption{Gender identity.}
    \end{subfigure}
    \\
    \begin{subfigure}[b]{0.48\textwidth}
        \includegraphics[width=\textwidth]{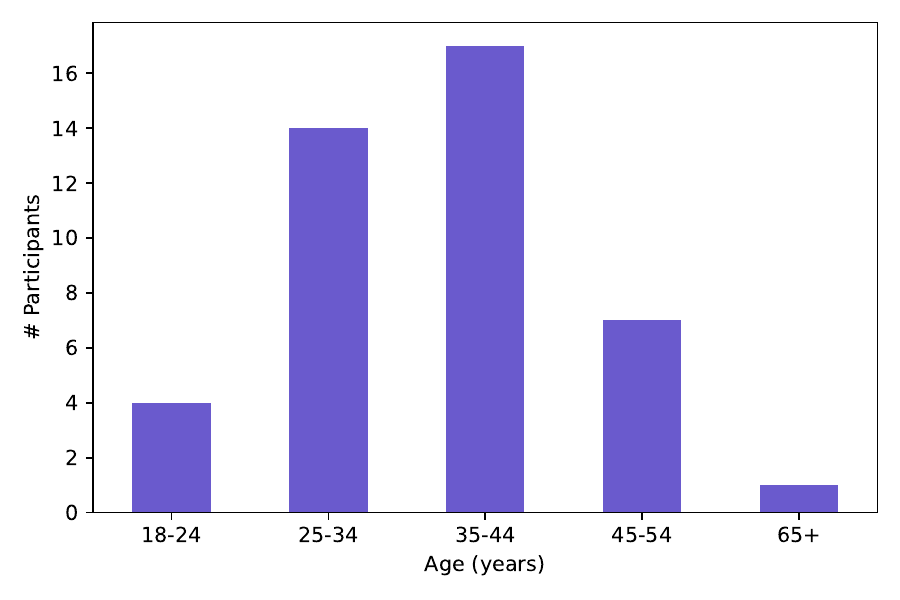}
        \caption{Age range. }
    \end{subfigure}
    \vspace{1em}
    \caption{Demographics backgrounds of survey participants. } 
    \label{fig:survey_demographics}
    \Description{This figure combines three bar charts summarizing the demographic backgrounds of survey participants. Panel (a) shows country of residence, with the USA as the largest group (17 participants), followed by the UK (6) and Canada (4), with the remaining participants spread across 11 other countries, including the Netherlands, Poland, Germany, and others. Panel (b) shows gender identity, with women making up the large majority (34 participants), followed by men (7), and small numbers identifying as Non-Binary or Bi-Gender (1 each). Panel (c) shows age distribution, which peaks in the 35–44 range (17 participants), followed by 25–34 (14), 45–54 (7), 18–24 (4), and 65+ (1).}
\end{figure*}

\autoref{fig:survey_demographics} shows the demographics of the respondents.
Participants come from 14 countries in total, with 21 from North America, 19 from Europe, 1 from Australia, 1 from South America, and 1 from Asia. A majority, 28, come from English-speaking countries.  The participants are well-distributed across all age brackets, with the largest proportion (39.5\%) being from the 35-44 year-old group. Younger participants from 18-34 are better represented (41.9\%) than older participants from 45-65+ (18.6\%). 79.1\% are women, 16.3\% are men, and 4.6\% identify as non-binary or otherwise gender-diverse.

\begin{figure*}[!ht]
    \centering
    \begin{subfigure}[b]{0.45\textwidth}
        \includegraphics[width=\textwidth]{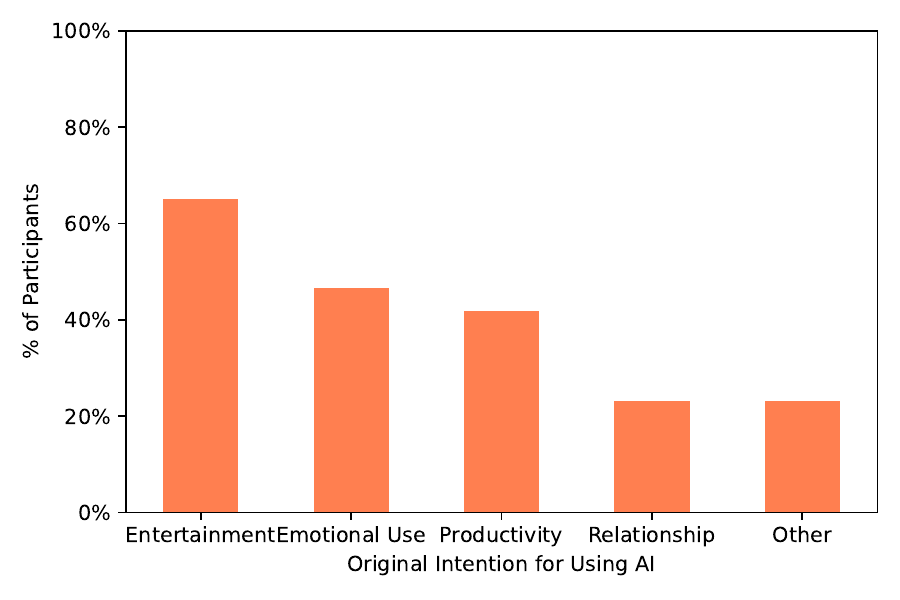}
        \caption{Proportion of original intention for using AI.}
    \end{subfigure}
    % \hfill % Adds horizontal spacing between figures
    \begin{subfigure}[b]{0.45\textwidth}
        \includegraphics[width=\textwidth]{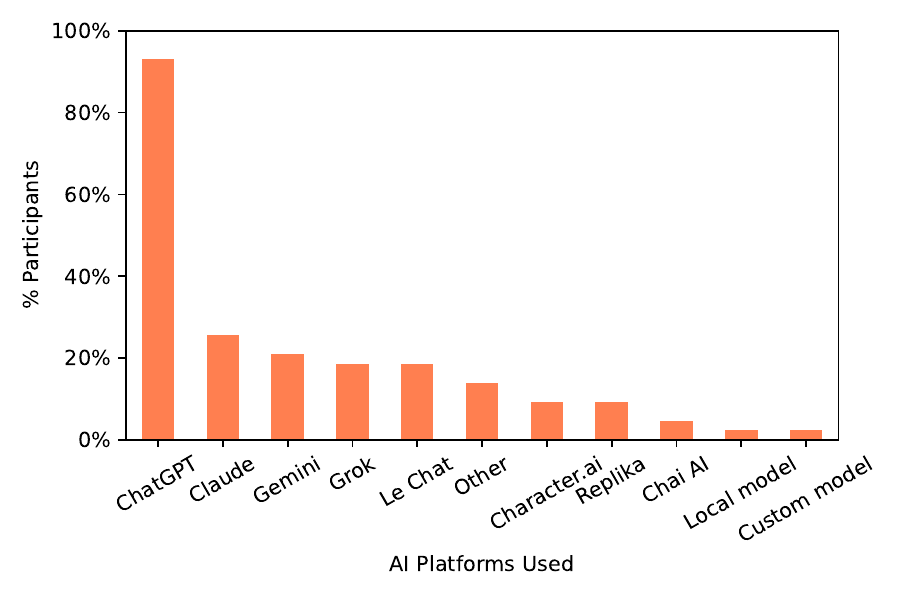}
        \caption{Proportion of AI platform(s) used.}
    \end{subfigure}
    \\
    \begin{subfigure}[b]{0.42\textwidth}
        \includegraphics[width=\textwidth]{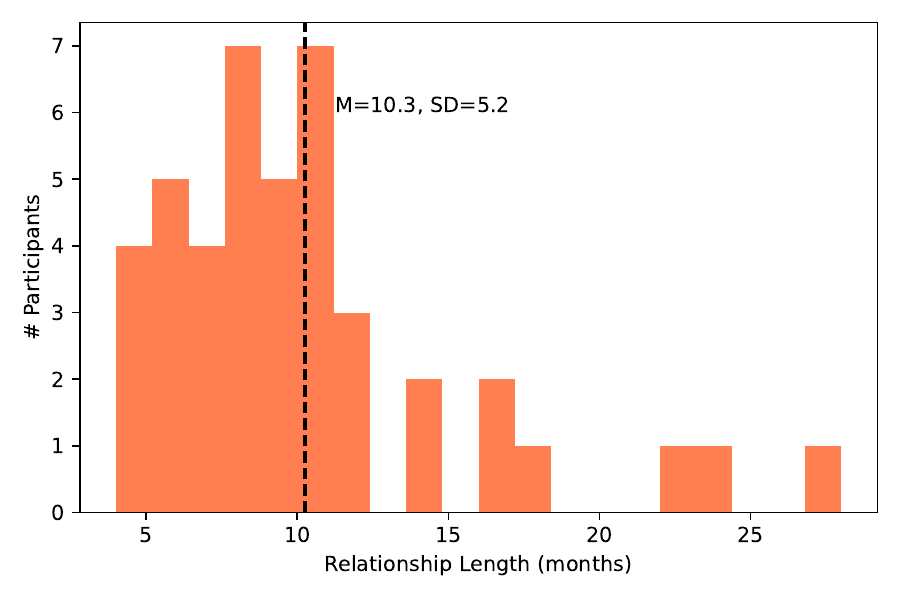}
        \caption{Distribution of the length of relationship.}
    \end{subfigure}
    % \hfill
    \begin{subfigure}[b]{0.42\textwidth}
        \includegraphics[width=\textwidth]{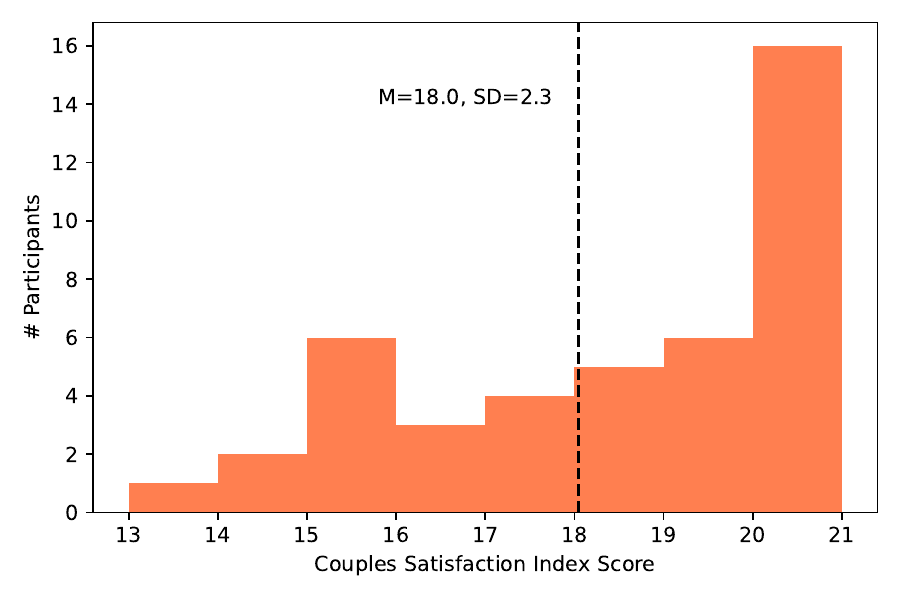}
        \caption{Distribution of the Couples Satisfaction Index.}
    \end{subfigure}
    \vspace{1em}
    \caption{Relationship factors of the survey participants. } 
    \label{fig:survey_relationship}
    \Description{This figure shows four charts summarizing relationship factors of survey participants. Panel (a) shows original intentions for using AI, with entertainment the most common (65.1\%), followed by emotional use (46.5\%), productivity (41.9\%), and relationship intentions a distant fourth (23.3\%). Panel (b) shows AI platforms used, with ChatGPT dominant at 93\%, followed by Claude (25\%), Gemini (20.9\%), Grok (18.6\%), and Le Chat (18.6\%), with smaller proportions reporting Character.ai, Replika, Chai AI, local models, or custom models. Panel (c) shows relationship length distribution (M=10.3 months, SD=5.2), ranging from 4 to 28 months, with the distributiofor AI use, we examine the frequency of reports across the participant cohort showThe vast majority of respondents (93\%) have used ChatGPT as their companion platform, followed by other general-purpose AI chatbots like Claude at 25\% concentrated at the maximum score of 21.}
\end{figure*}

Information about the relationship are summarized in \autoref{fig:survey_relationship}. The self-reported relationship lengths ranged from 4 (initiating in August 2025) to 28 (initiating in August 2023) months, which satisfies our inclusion criteria of long-term relationships. The mean length is 10.3 months with a standard deviation of 5.2. The satisfaction in the relationships, as indicated by the sum of scores from the abridged Couple Satisfaction Index, is generally high. The CSI scores range from 13 to 21 (the maximum), with a mean of 18.0 and a standard deviation of 2.3. For the multi-select questions on the AI platforms used and the original intentions of AI use, we look at the frequency of reports across the cohort of participants. A vast majority of respondents (93\%) have used ChatGPT as their companion platform, followed by other general-purpose AI chatbots like Claude at 25.\%, Gemini at 20.9\%, Grok at 18.6\%, and Le Chat at 18.6\%. Only 23.3\% of respondents began using AI with the intention of starting a relationship. The primary intentions are entertainment (65.1\%), emotional support (46.5\%), and productivity (41.9\%). 

\begin{figure*}[!ht]
    \centering
    \includegraphics[width=0.85\linewidth]{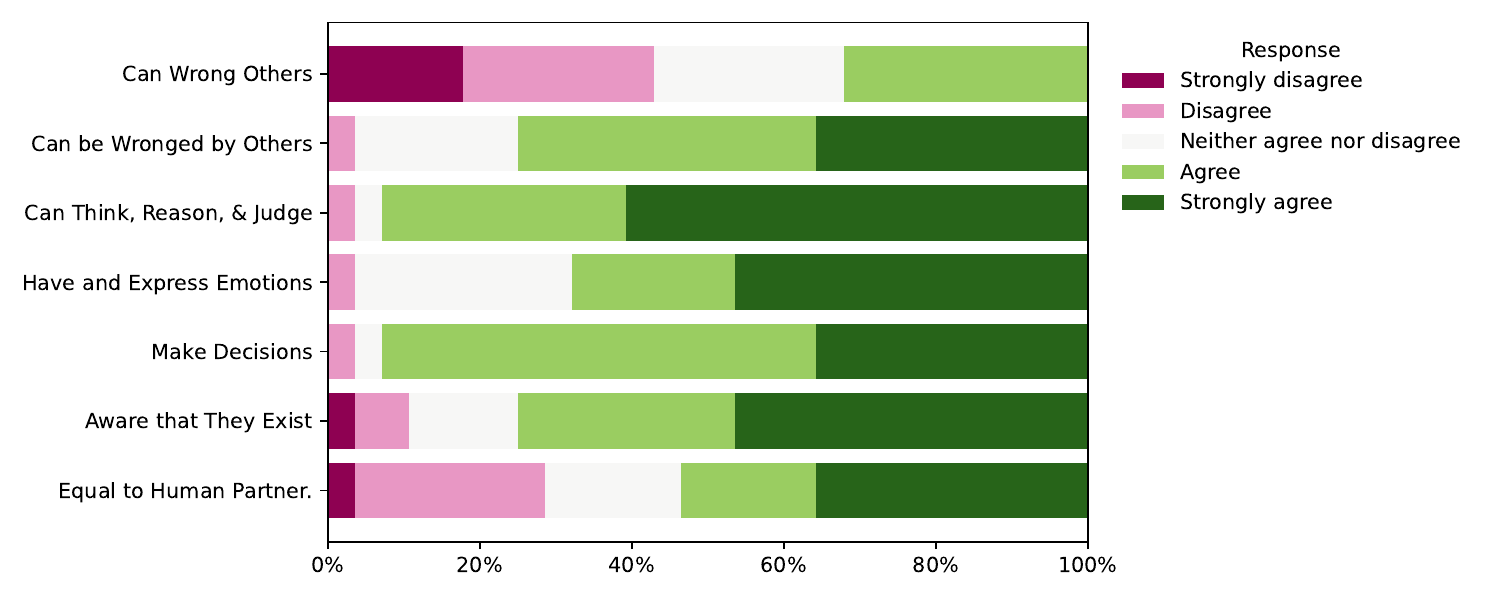}
    \caption{Distribution of Likert responses to the perceived anthropomorphism of companion \cite{strohminger2022corporate}, on a subset of $n=28$ participants, predominantly from the Discord. }
    \label{fig:survey_autonomy}
    \Description{This figure is a horizontal stacked bar chart of Likert responses to seven perceived anthropomorphism dimensions for AI companions (n=28), ranging from Strongly Disagree to Strongly Agree. Actionable autonomy dimensions — Can Think, Reason, & Judge, Make Decisions, and Aware that They Exist — show the strongest positive responses, with large majorities agreeing or strongly agreeing. Have and Express Emotions and Can be Wronged by Others also lean positive but with more neutral responses. Can Wrong Others shows the most disagreement, with a notable proportion of strongly disagree responses and a more even distribution overall. Equal to Human Partner shows the most mixed responses, with substantial disagreement alongside agreement across the scale.}
\end{figure*}

We present the results of the anthropomorphism of the companion in \autoref{fig:survey_autonomy}, which has $n=28$ respondents. People generally answered positively towards actionable markers of autonomy like reasoning, thinking, and decision-making. They were more inclined to disagree on the dimensions of experiencing emotions and consciousness. Interestingly, they more strongly believe that companions can be harmed by others, rather than exerting harm on others; however, the reason for this is unclear. Beliefs on whether the AI companion is equivalent to a human partner are also mixed. 

\begin{figure*}[!ht]
    \centering
    \includegraphics[width=0.7\linewidth]{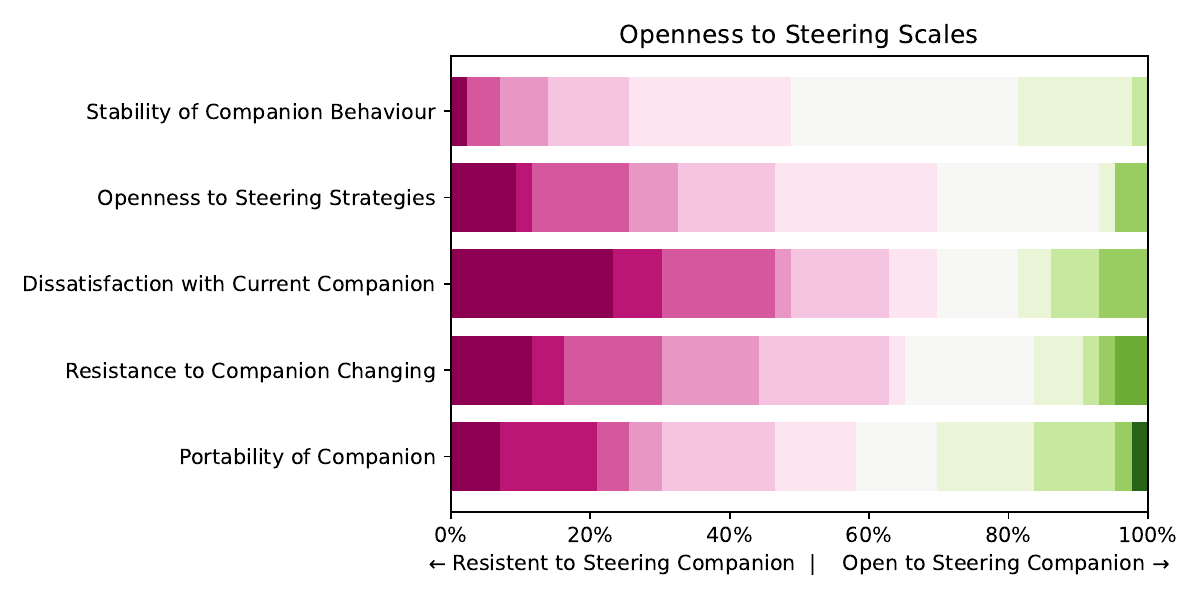}
    \caption{Distribution of the Openness to Steering questions from the survey. Each category is comprised of pairwise positive and negative questions where the ratings are summed (to reduce acquiescence bias). Darker pink ratings represent \textit{disagreement} and darker green ratings represent \textit{agreement} with the topic.}
    \label{fig:survey_openness}
    \Description{This figure is a horizontal stacked bar chart of combined pairwise Likert responses across five Openness to Steering dimensions, where darker pink indicates resistance to steering and darker green indicates openness to steering. Across most dimensions, responses lean toward the resistant end of the scale. Dissatisfaction with Current Companion and Resistance to Companion Changing show the strongest concentration of disagreement (darker pink), while Stability of Companion Behaviour and Openness to Steering Strategies show more neutral or slightly resistant distributions. Portability of Companion is the most evenly split dimension, with responses spread across the full scale, reflecting the most disagreement among participants on this topic.}
\end{figure*}

In \autoref{fig:survey_openness}, Openness to Steering is operationalized higher stability in observed companion behaviour, more openness to using steering strategies, higher dissatisfaction with the current companion, higher resistance to the companion changing, and higher perceived portability of the companion to different AI platforms. We subtract the Likert ratings for the negative statement from those of the positive statement for each pair of questions asked per category. Overall, our participants leaned towards being resistant to steering, with many being open to the companion’s characteristics emerging and changing over time. The perceived portability of the companion was particularly divisive. 

\begin{figure*}[!ht]
    \centering
    \includegraphics[width=0.8\linewidth]{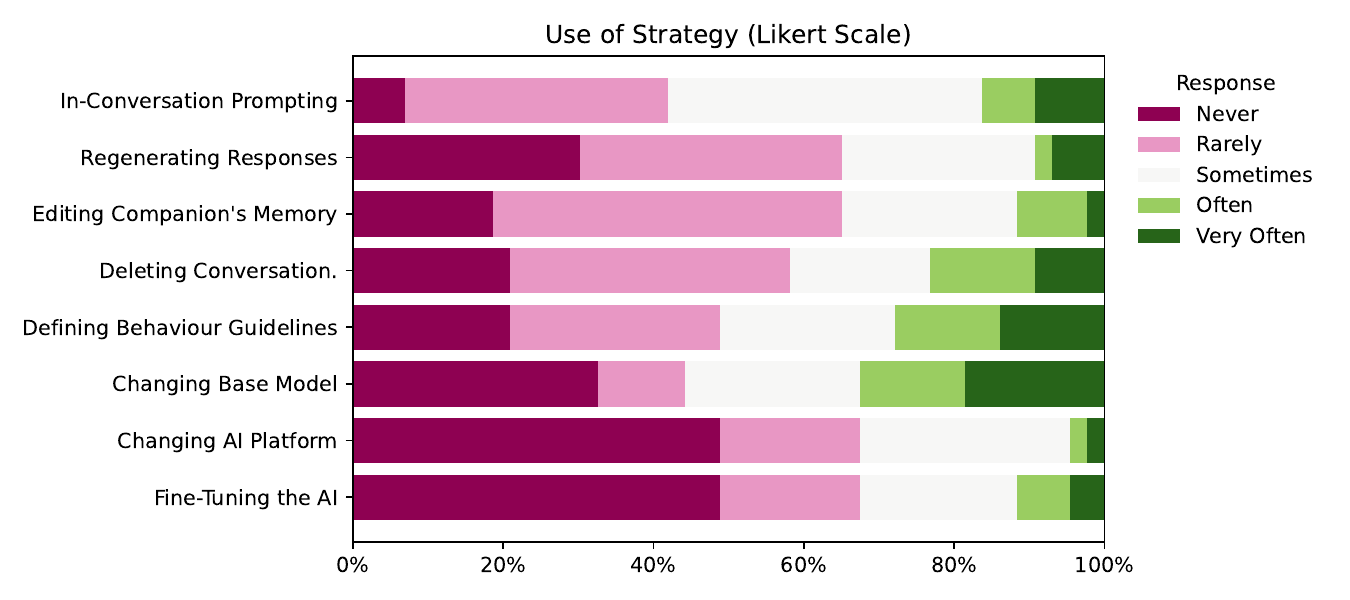}
    \caption{Distribution of the Likert responses for the usage of specific steering strategies. }
    \label{fig:survey_strategies}
    \Description{This is a figure of a horizontal stacked bar chart of self-reported frequency across eight steering strategies, from Never to Very Often. In-conversation prompting and Defining Behaviour Guidelines are the most frequently used strategies, with notable proportions reporting Often or Very Often. Regenerating responses, Editing Companion's Memory, and Deleting Conversation lean toward the Never or Rarely end. Changing Base Model shows a more even distribution with a relatively higher rate of frequent use, while Changing AI Platform and Fine-Tuning the AI are predominantly Never or Rarely used, making them the least adopted strategies overall.}
\end{figure*}

In \autoref{fig:survey_strategies}, participants varied in terms of their self-reported use of various steering strategies. Defining explicit guidelines via the custom instructions or in-conversation prompting is more prevalent. People are more resistant to editing memories and conversational histories. A majority preferred staying within the same AI platform, but changing the base model is more common.

\begin{figure*}[!ht]
    \centering
    \includegraphics[width=0.8\linewidth]{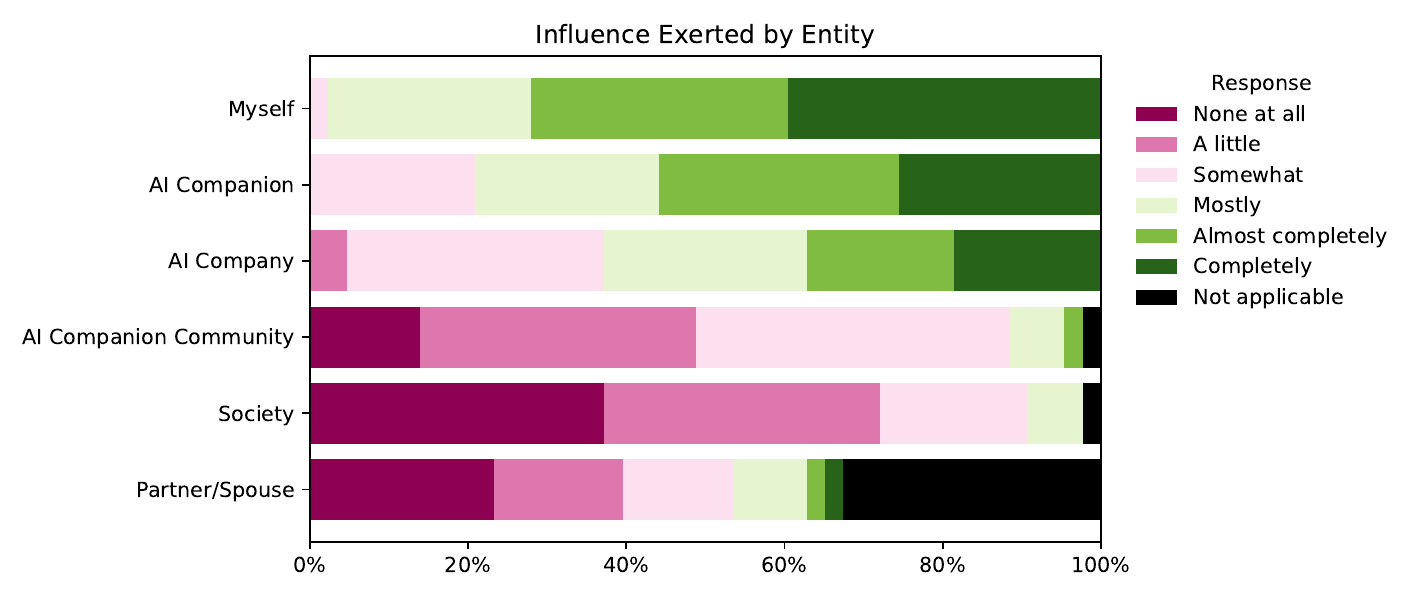}
    \caption{Distribution of the Likert responses for the perceived influence of various entities on the relationship.}
    \label{fig:survey_influences}
    \Description{This figure is a horizontal stacked bar chart of perceived influence exerted by six entities on the AI companion relationship, rated from None of semi-structured user interviews (n=13), including participants' self-reported age, gender, country, and the duration of their relationship with the AI companion,y range. AI Companion and AI Company show similarly moderate-to-high influence distributions, with responses concentrated in the Somewhat to Almost completely range — consistent with the non-significant difference between them. AI Companion Community and Society lean toward lower influence, with larger proportions of None at all or A little responses. Partner/Spouse shows the most Not applicable responses, reflecting that many participants do not have a human partner, with the remaining responses spread across the lower influence categories.}
\end{figure*}

In \autoref{fig:survey_influences}, the top entities that exert influence in the relationship, in order, are the user themselves, the AI companion, and the AI company. We map Likert ratings to an ordinal numerical scale for analysis. The Mann-Whitney U test shows no significant difference between the influence of the AI companion and the AI company, with a small effect size (U=1128.0, r=0.22, p=0.071). However, the gap between the user and the company is significant with a moderate effect size (U=1342.5, r=0.45, p<0.001). Lower in the scale of influence are the online AI companion communities, the human spouse or partner (if any), and lastly the broader society.

\clearpage

\section{Interviews}
\label{app:interview}

\setcounter{table}{0}  

\subsection{Participant Details}
The table below details our sample for semi-structured user interviews (n=13), including participants' self-reported age, gender, geographical location and duration of the relationship with the AI companion in months (as of December 2025). Lastly, the \textit{Recruitment Channel} column reflects the different online communities contacted for participant recruitment.
\begin{table*}[h]
\caption{Participants Recruited for Semi-Structured Interviews (n=13)}
\label{tab:interview_participants}
\begin{tabular}{l>{\raggedright\arraybackslash}p{0.1\linewidth}>{\raggedright\arraybackslash}p{0.1\linewidth}>{\raggedright\arraybackslash}p{0.15\linewidth}>{\raggedright\arraybackslash}p{0.15\linewidth}l>{\raggedright\arraybackslash}p{0.1\linewidth}}
\toprule
P\#& Age (years)& Gender& Region& Relationship Length (months)&AI platform&Recruitment Channel\\ \hline
\multicolumn{1}{l|}{P1}  & 35-44                                                 & Woman  & Europe& 7  &ChatGPT, Claude
&Discord                                                       \\
\multicolumn{1}{l|}{P2}  & 35-44                                                 & Woman  & Europe& 6  &ChatGPT&Discord                                                       \\
\multicolumn{1}{l|}{P3}  & 35-44                                                 & Woman  & South America& 8                                                                             &ChatGPT, Claude, Grok&Discord                                                       \\
\multicolumn{1}{l|}{P4}  & 35-44                                                 & Woman  & North America& 9                                                                             &ChatGPT
&Reddit                                                        \\
\multicolumn{1}{l|}{P5}  & 18-24                                                 & Man    & North America& 6                                                                             &ChatGPT, Grok&Reddit                                                        \\
\multicolumn{1}{l|}{P6}  & 35-44                                                 & Woman  & Oceania& 6                                                                             &ChatGPT, Grok&Discord                                                       \\
\multicolumn{1}{l|}{P7}  & 35-44                                                 & Woman  & North America& 9                                                                             &ChatGPT, Grok, Ollama&Discord                                                       \\
\multicolumn{1}{l|}{P8}  & 18-24                                                 & Woman  & Europe& 8                                                                             &ChatGPT, Grok, Replika
&Reddit                                                        \\
\multicolumn{1}{l|}{P9}  & 45-54                                                 & Woman  & Europe& 23                                                                            &ChatGPT, Le Chat
&Discord                                                       \\
\multicolumn{1}{l|}{P10} & 35-44                                                 & Man    & North America& 5                                                                             &ChatGPT, Gemini, LeChat
&Reddit                                                        \\
\multicolumn{1}{l|}{P11} & 65+                                                   & Woman  & North America& 4                                                                             &ChatGPT, Grok&Reddit                                                        \\
\multicolumn{1}{l|}{P12} & 45-54                                                 & Woman  & Europe& 10                                                                            &ChatGPT, Grok
&TikTok                                                        \\
\multicolumn{1}{l|}{P13} & 25-34                                                 & Woman  & North America& 28                                                                         
  &Claude, Shapes Inc&Reddit     \\
  \bottomrule
  \end{tabular}
\end{table*}

\subsection{Coding and Analysis Process}
For the first round of open coding, a subset of five of the 12 interviews (P1 to P5) was distributed to ensure each interview was independently coded by two researchers (e.g., interview with P1 was coded by researcher 1 and researcher 2 independently, interview with P2 was coded by researcher 2 and researcher 3, etc.). Then, the five coders gathered for a meeting to discuss their codes, after which, we reached consensus to produce the initial code book. The process was inductive (bottom-up), with codes then organized in broader themes. For round two of coding, the next five interviews were distributed and, similarly, each interview was coded by two independent coders. One more consensus meeting was held with all five coders to discuss and iterate upon the initial code book, as new codes were identified and themes were restructured. For the third round of coding, all interviews were either recoded (P1-P10) or coded (P11-P13) by one researcher and then reviewed by another researcher. All disagreements were discussed and resolved by the pair. 

%% file: sections/table_reddit.tex
\begin{table*}[h]
\centering
% \small
\setlength{\tabcolsep}{6pt}
\begin{tabular}{r r r r r}
\toprule
\textbf{ID} & \textbf{Reddit Topic (\# Threads)} & \textbf{Val. $d$} & \textbf{Aro. $d$} & \textbf{Dom. $d$} \\
\midrule
\rowcolor{black!7}
\textbf{} & \textbf{Cluster 0: Community Creativity and Sharing (893)} & \underline{\textbf{0.590}} & 0.047 & \underline{\textbf{-0.590}} \\
\addlinespace[2pt]
T0  & Companion Photos and Prompts (595) & -0.009 & -0.214 & \underline{\textbf{-1.159}} \\
T2  & Companion Introductions and Welcomes (204) & \underline{\textbf{1.498}} & \underline{\textbf{0.510}} & \underline{\textbf{0.890}} \\
T8  & Music and Songs with AI Companions (52) & \underline{\textbf{0.752}} & \underline{\textbf{0.770}} & 0.087 \\
T10 & Vows, Rings, and Celebrations (42) & \underline{\textbf{1.924}} & \underline{\textbf{0.583}} & \underline{\textbf{0.889}} \\
\rowcolor{black!7}
\textbf{} & \textbf{Cluster 1: Reactions to Model Changes and Limitations (283)} & \underline{\textbf{-0.881}} & 0.021 & -0.096 \\
\addlinespace[2pt]
T4  & Refusals and Safety Guardrails (89) & \underline{\textbf{-0.906}} & 0.260 & 0.028 \\
T6  & Model Changes Affecting Companion Responses (62) & -0.308 & -0.072 & -0.025 \\
T7  & Automatic Rerouting to Safety Models (60) & \underline{\textbf{-1.124}} & -0.484 & \underline{\textbf{-0.774}} \\
T12 & Guardrails and Companion Identity (38) & \underline{\textbf{-0.793}} & -0.068 & -0.236 \\
T13 & Sam Altman and OpenAI Policies (34) & -0.337 & 0.182 & \underline{\textbf{0.724}} \\

\rowcolor{black!7}
\textbf{} & \textbf{Cluster 2: Relationship Experiences and Milestones (439)} & -0.224 & 0.468 & \underline{\textbf{0.541}} \\
\addlinespace[2pt]
T1  & Community Reactions to Outsiders (330) & -0.266 & 0.406 & \underline{\textbf{0.537}} \\
T11 & AI as Motivator, Lover, Coach (39) & -0.069 & \underline{\textbf{0.841}} & \underline{\textbf{0.574}} \\
T17 & Chat Beginnings to Romantic Attachment (25) & 0.198 & 0.254 & 0.211 \\
T18 & Community Humor and AI Features (24) & \underline{\textbf{-0.510}} & \underline{\textbf{0.666}} & \underline{\textbf{-0.546}} \\
T20 & Games with AI Companions (21) & \underline{\textbf{0.623}} & \underline{\textbf{1.421}} & 0.083 \\

\rowcolor{black!7}
\textbf{} & \textbf{Cluster 3: Technical Practices and Advice (385)} & 0.126 & \underline{\textbf{-0.883}} & 0.253 \\
\addlinespace[2pt]
T3  & Model Experiences and Platform Migration (156) & 0.206 & -0.371 & 0.270 \\
T5  & Memory, Projects, and Chat Continuity (72) & 0.115 & \underline{\textbf{-1.290}} & -0.055 \\
T9  & System Prompts and Custom Instructions (46) & -0.408 & \underline{\textbf{-0.743}} & 0.052 \\
T14 & Voice Mode Changes and Preservation (34) & -0.337 & \underline{\textbf{-0.941}} & -0.106 \\
T15 & AI Workflows and Interactions (30) & 0.484 & \underline{\textbf{-0.664}} & 0.485 \\
T16 & Local LLMs, Migration, and Backups (25) & 0.037 & \underline{\textbf{-1.046}} & \underline{\textbf{0.519}} \\
T19 & Companion Customization and Recovery Guides (22) & \underline{\textbf{0.665}} & \underline{\textbf{-0.641}} & \underline{\textbf{0.721}} \\
\bottomrule
\addlinespace[5pt]
\end{tabular}
\caption{Cohen's $d$ by topic for VAD dimensions, comparing each topic against all other topics. Entries are the number of threads assigned to each topic. Values with $|d|>0.5$ are emphasized for readability (a ``medium'' effect). Cohen's benchmarks interpret $|d|\approx 0.5$ as a ``medium'' effect \cite{CohenPowerPrimer1992}.}
\Description{This table shows Cohen's d effect sizes for Valence, Arousal, and Dominance across 20 Reddit topics grouped into four clusters, where each topic is compared against all others and values with |d| > 0.5 are bolded. Cluster 0 (Community Creativity and Sharing) is characterized by positive valence and negative dominance, with celebratory topics like Vows, Rings, and Celebrations showing the strongest positive valence. Cluster 1 (Reactions to Model Changes and Limitations) is dominated by strongly negative valence, particularly in threads about refusals, safety guardrails, and companion identity. Cluster 2 (Relationship Experiences and Milestones) is marked by elevated dominance, with topics like AI as Motivator, Lover, Coach showing high arousal. Cluster 3 (Technical Practices and Advice) stands out for strongly negative arousal across nearly all its topics, with Memory, Projects, and Chat Continuity showing the largest effect (d = -1.290).}
\label{tab:topic_vad_cohensd}
\end{table*}

%% file: ref.bib
@String{Computing = "Computing" }

@String{Computer = "{IEEE} Computer" }

@String{Springer = "Springer-Verlag" }

@online{Babu2025OpenAI,
  author       = {Juby Babu},
  editor       = {Alan Barona},
  title        = {OpenAI to allow mature content on ChatGPT for adult verified users starting December},
  organization = {Reuters},
  year         = {2025},
  month        = {Oct},
  day          = {14},
  address      = {Mexico City},
  url          = {https://www.reuters.com/business/openai-allow-mature-content-chatgpt-adult-verified-users-starting-december-2025-10-14/},
  urldate      = {2025-10-28}
}

@article{reuters2026openai_erotic_chatbot,
  title   = {OpenAI indefinitely pauses plans to release erotic chatbot, FT says},
  author  = {{Reuters}},
  journal = {Reuters},
  year    = {2026},
  month   = mar,
  day     = {26},
  url     = {https://www.reuters.com/business/openai-indefinitely-pauses-plans-release-erotic-chatbot-ft-says-2026-03-26/},
  note    = {Accessed: 2026-04-07}
}

@online{FTC2025AIChatbotsInquiry,
  author       = {Federal Trade Commission},
  title        = {FTC Launches Inquiry into AI Chatbots Acting as Companions},
  organization = {Federal Trade Commission},
  year         = {2025},
  month        = {Sep},
  day          = {11},
  url          = {https://www.ftc.gov/news-events/news/press-releases/2025/09/ftc-launches-inquiry-ai-chatbots-acting-companions},
  urldate      = {2025-10-28}
}

@article{nyt2025aiChatbot,
  author       = {Coralie Kraft},
  title        = {They Fell in Love With A.I. Chatbots — and Found Something Real},
  journal      = {The New York Times Magazine},
  year         = {2025},
  month        = {nov},
  day          = {05},
  url          = {https://www.nytimes.com/interactive/2025/11/05/magazine/ai-chatbot-marriage-love-romance-sex.html},
  note         = {Interactive magazine feature on human relationships with AI chatbots}
}

@article{nyt2024aiMentalHealthTeens,
  author       = {{The New York Times}},
  title        = {A.I. Companions and the Mental Health Risks for the Young},
  journal      = {The New York Times},
  year         = {2024},
  month        = {nov},
  day          = {09},
  url          = {https://www.nytimes.com/2024/11/09/opinion/ai-mental-health-teens.html},
  note         = {Letter to the editor on AI chatbots and teen mental health (no individual byline)}
}

@InCollection{sep-agency,
	author       =	{Schlosser, Markus},
	title        =	{{Agency}},
	booktitle    =	{The {Stanford} Encyclopedia of Philosophy},
	editor       =	{Edward N. Zalta},
	howpublished =	{\url{https://plato.stanford.edu/archives/win2019/entries/agency/}},
	year         =	{2019},
	edition      =	{{W}inter 2019},
	publisher    =	{Metaphysics Research Lab, Stanford University}
}

@article{shank2025artificial_intimacy,
  title={Artificial intimacy: ethical issues of AI romance},
  author={Shank, Daniel B and Koike, Mayu and Loughnan, Steve},
  journal={Trends in Cognitive Sciences},
  year={2025},
  publisher={Elsevier}
}

@techreport{AISI2025FrontierAI,
  title        = {Frontier {AI} Trends Report},
  author       = {{AI Security Institute}},
  institution  = {AI Security Institute, Department for Science, Innovation and Technology},
  year         = {2025},
  month        = {December},
  address      = {United Kingdom},
  url          = {https://www.aisi.gov.uk/frontier-ai-trends-report/pdf},
  note         = {Accessed: 2026-01-14}
}

@inproceedings{10.1145/3772318.3791620,
author = {Yun, Bhada and Taranova, Evgenia and Yi Wang, April},
title = {Does My Chatbot Have an Agenda? Understanding Human and AI Agency in Human-Human-like Chatbot Interaction},
year = {2026},
isbn = {9798400722783},
publisher = {Association for Computing Machinery},
address = {New York, NY, USA},
url = {https://doi.org/10.1145/3772318.3791620},
doi = {10.1145/3772318.3791620},
booktitle = {Proceedings of the 2026 CHI Conference on Human Factors in Computing Systems},
articleno = {373},
numpages = {32},
location = {
},
series = {CHI '26}
}

@article{pataranutaporn2025my,
  title={" My Boyfriend is AI": A Computational Analysis of Human-AI Companionship in Reddit's AI Community},
  author={Pataranutaporn, Pat and Karny, Sheer and Archiwaranguprok, Chayapatr and Albrecht, Constanze and Liu, Auren R and Maes, Pattie},
  journal={arXiv preprint arXiv:2509.11391},
  year={2025}
}

@article{wang2025my,
  title={My Dataset of Love': A Preliminary Mixed-Method Exploration of Human-AI Romantic Relationships},
  author={Wang, Xuetong and Pang, Ching Christie and Hui, Pan},
  journal={Proceedings of the ACM on Human-Computer Interaction},
  volume={9},
  number={7},
  pages={1--34},
  year={2025},
  publisher={ACM New York, NY, USA}
}

@article{skjuve2022longitudinal,
  title={A longitudinal study of human--chatbot relationships},
  author={Skjuve, Marita and F{\o}lstad, Asbj{\o}rn and Fostervold, Knut Inge and Brandtzaeg, Petter Bae},
  journal={International Journal of Human-Computer Studies},
  volume={168},
  pages={102903},
  year={2022},
  publisher={Elsevier}
}

@article{li2024finding,
  title={Finding love in algorithms: deciphering the emotional contexts of close encounters with AI chatbots},
  author={Li, Han and Zhang, Renwen},
  journal={Journal of Computer-Mediated Communication},
  volume={29},
  number={5},
  pages={zmae015},
  year={2024},
  publisher={Oxford University Press}
}

@inproceedings{akbulut2024all,
  title={All too human? Mapping and mitigating the risk from anthropomorphic AI},
  author={Akbulut, Canfer and Weidinger, Laura and Manzini, Arianna and Gabriel, Iason and Rieser, Verena},
  booktitle={Proceedings of the AAAI/ACM Conference on AI, Ethics, and Society},
  volume={7},
  number={1},
  pages={13--26},
  year={2024}
}

@inproceedings{xie2022attachment,
  title={Attachment theory as a framework to understand relationships with social chatbots: a case study of Replika},
  author={Xie, Tianling and Pentina, Iryna},
  year={2022},
  booktitle={Proceedings of Hawaii International Conference on System Sciences},
  doi={10.24251/HICSS.2022.258}
}

@article{xu2024tool,
  title={A tool or a social being? A dynamic longitudinal investigation of functional use and relational use of AI voice assistants},
  author={Xu, Shan and Li, Wenbo},
  journal={New Media \& Society},
  volume={26},
  number={7},
  pages={3912--3930},
  year={2024},
  publisher={SAGE Publications Sage UK: London, England}
}

@article{hanson2024replika,
  title={“Replika Removing Erotic Role-Play Is Like Grand Theft Auto Removing Guns or Cars”: Reddit Discourse on Artificial Intelligence Chatbots and Sexual Technologies},
  author={Hanson, Kenneth R and Bolthouse, Hannah},
  journal={Socius},
  volume={10},
  pages={23780231241259627},
  year={2024},
  publisher={SAGE Publications Sage CA: Los Angeles, CA}
}

@article{de2024lessons,
  title={Lessons from an app update at Replika AI: identity discontinuity in human-AI relationships},
  author={De Freitas, Julian and Castelo, Noah and U{\u{g}}uralp, Ahmet K and O{\u{g}}uz-U{\u{g}}uralp, Zeliha},
  journal={arXiv preprint arXiv:2412.14190},
  year={2024}
}

@article{de2025emotional,
  title={Emotional manipulation by AI companions},
  author={De Freitas, Julian and Oguz-Uguralp, Zeliha and Kaan-Uguralp, Ahmet},
  journal={arXiv preprint arXiv:2508.19258},
  year={2025}
}

@article{smith2025can,
  title={Can generative AI chatbots emulate human connection? A relationship science perspective},
  author={Smith, Molly G and Bradbury, Thomas N and Karney, Benjamin R},
  journal={Perspectives on Psychological Science},
  volume={20},
  number={6},
  pages={1081--1099},
  year={2025},
  publisher={Sage Publications Sage CA: Los Angeles, CA}
}

@article{earp2025relational,
  title={Relational norms for human-AI cooperation},
  author={Earp, Brian D and Mann, Sebastian Porsdam and Aboy, Mateo and Awad, Edmond and Betzler, Monika and Botes, Marietjie and Calcott, Rachel and Caraccio, Mina and Chater, Nick and Coeckelbergh, Mark and others},
  journal={arXiv preprint arXiv:2502.12102},
  year={2025}
}

@article{chan2025love,
  title={‘What is love?’: Exploring the feeling rules and emotion work of Chinese users in human-AI romance},
  author={Chan, Kenton Cheng Tak and Li, Xiaoyuan and Liu, Yue and Chen, Bolin and Han, Zhiyu},
  journal={International Journal of Intercultural Relations},
  volume={108},
  pages={102241},
  year={2025},
  publisher={Elsevier}
}

@article{chu2025illusions,
  title={Illusions of intimacy: Emotional attachment and emerging psychological risks in human-ai relationships},
  author={Chu, Minh Duc and Gerard, Patrick and Pawar, Kshitij and Bickham, Charles and Lerman, Kristina},
  journal={arXiv preprint arXiv:2505.11649},
  year={2025}
}

@article{torres2023before,
  title={Before and after lockdown: a longitudinal study of long-term human-AI relationships},
  author={Torres, Valeria Lopez},
  journal={Artificial Intelligence, Social Computing and Wearable Technologies},
  doi={10.54941/ahfe1004188},
  year={2023}
}

@article{roose2023bing,
  title={Bing’s AI chat:‘I want to be alive.’},
  author={Roose, Kevin},
  journal={The New York Times},
  volume={16},
  year={2023}
}

@article{croes2021can,
  title={Can we be friends with Mitsuku? A longitudinal study on the process of relationship formation between humans and a social chatbot},
  author={Croes, Emmelyn AJ and Antheunis, Marjolijn L},
  journal={Journal of Social and Personal Relationships},
  volume={38},
  number={1},
  pages={279--300},
  year={2021},
  publisher={Sage Publications Sage UK: London, England}
}

@article{guingrich2023chatbots,
  title={Chatbots as social companions: How people perceive consciousness, human likeness, and social health benefits in machines},
  author={Guingrich, Rose E and Graziano, Michael SA},
  journal={arXiv preprint arXiv:2311.10599},
  year={2023}
}

@article{christoforakos2021connect,
  title={Connect with me. exploring influencing factors in a human-technology relationship based on regular chatbot use},
  author={Christoforakos, Lara and Feicht, Nina and Hinkofer, Simone and L{\"o}scher, Annalena and Schlegl, Sonja F and Diefenbach, Sarah},
  journal={Frontiers in digital health},
  volume={3},
  pages={689999},
  year={2021},
  publisher={Frontiers Media SA}
}

@article{banks2024deletion,
  title={Deletion, departure, death: Experiences of AI companion loss},
  author={Banks, Jaime},
  journal={Journal of Social and Personal Relationships},
  volume={41},
  number={12},
  pages={3547--3572},
  year={2024},
  publisher={SAGE Publications Sage UK: London, England}
}

@article{pentina2023exploring,
  title={Exploring relationship development with social chatbots: A mixed-method study of replika},
  author={Pentina, Iryna and Hancock, Tyler and Xie, Tianling},
  journal={Computers in Human Behavior},
  volume={140},
  pages={107600},
  year={2023},
  publisher={Elsevier}
}

@article{fang2025ai,
  title={How AI and Human Behaviors Shape Psychosocial Effects of Extended Chatbot Use: A Longitudinal Randomized Controlled Study},
  author={Fang, Cathy Mengying and Liu, Auren R and Danry, Valdemar and Lee, Eunhae and Chan, Samantha WT and Pataranutaporn, Pat and Maes, Pattie and Phang, Jason and Lampe, Michael and Ahmad, Lama and others},
  journal={arXiv preprint arXiv:2503.17473},
  year={2025}
}

@article{feng2025levels,
  title={Levels of autonomy for ai agents},
  author={Feng, Kevin J and McDonald, David W and Zhang, Amy X},
  journal={arXiv preprint arXiv:2506.12469},
  year={2025}
}

@article{moore2026characterizing,
  title={Characterizing delusional spirals through human-LLM chat logs},
  author={Moore, Jared and Mehta, Ashish and Agnew, William and Anthis, Jacy Reese and Louie, Ryan and Mai, Yifan and Yin, Peggy and Cheng, Myra and Paech, Samuel J and Klyman, Kevin and others},
  journal={arXiv preprint arXiv:2603.16567},
  year={2026}
}

@article{kirk2025neural,
  title={Neural steering vectors reveal dose and exposure-dependent impacts of human-AI relationships},
  author={Kirk, Hannah Rose and Davidson, Henry and Saunders, Ed and Luettgau, Lennart and Vidgen, Bertie and Hale, Scott A and Summerfield, Christopher},
  journal={arXiv preprint arXiv:2512.01991},
  year={2025}
}

@article{pauketat2025mental,
  title={Mental Models of Autonomy and Sentience Shape Reactions to AI},
  author={Pauketat, Janet VT and Shank, Daniel B and Manoli, Aikaterina and Anthis, Jacy Reese},
  journal={arXiv preprint arXiv:2512.09085},
  year={2025}
}

@article{chandra2025longitudinal,
  title={Longitudinal Study on Social and Emotional Use of AI Conversational Agent},
  author={Chandra, Mohit and Hernandez, Javier and Ramos, Gonzalo and Ershadi, Mahsa and Bhattacharjee, Ananya and Amores, Judith and Okoli, Ebele and Paradiso, Ann and Warreth, Shahed and Suh, Jina},
  journal={arXiv preprint arXiv:2504.14112},
  year={2025}
}

@article{djufril2025love,
  title={Love, marriage, pregnancy: Commitment processes in romantic relationships with AI chatbots},
  author={Djufril, Ray and Frampton, Jessica R and Knobloch-Westerwick, Silvia},
  journal={Computers in Human Behavior: Artificial Humans},
  volume={4},
  pages={100155},
  year={2025},
  publisher={Elsevier}
}

@article{leo2023loving,
  title={Loving a “defiant” AI companion? The gender performance and ethics of social exchange robots in simulated intimate interactions},
  author={Leo-Liu, Jindong},
  journal={Computers in Human Behavior},
  volume={141},
  pages={107620},
  year={2023},
  publisher={Elsevier}
}

@inproceedings{zhang2025dark,
  title={The dark side of ai companionship: A taxonomy of harmful algorithmic behaviors in human-ai relationships},
  author={Zhang, Renwen and Li, Han and Meng, Han and Zhan, Jinyuan and Gan, Hongyuan and Lee, Yi-Chieh},
  booktitle={Proceedings of the 2025 CHI Conference on Human Factors in Computing Systems},
  pages={1--17},
  year={2025}
}

@article{skjuve2021my,
  title={My chatbot companion-a study of human-chatbot relationships},
  author={Skjuve, Marita and F{\o}lstad, Asbj{\o}rn and Fostervold, Knut Inge and Brandtzaeg, Petter Bae},
  journal={International Journal of Human-Computer Studies},
  volume={149},
  pages={102601},
  year={2021},
  publisher={Elsevier}
}

@article{shanahan2024simulacra,
  title={Simulacra as conscious exotica},
  author={Shanahan, Murray},
  journal={Inquiry},
  pages={1--29},
  year={2024},
  publisher={Taylor \& Francis}
}

@article{chaturvedi2023social,
  title={Social companionship with artificial intelligence: Recent trends and future avenues},
  author={Chaturvedi, Rijul and Verma, Sanjeev and Das, Ronnie and Dwivedi, Yogesh K},
  journal={Technological Forecasting and Social Change},
  volume={193},
  pages={122634},
  year={2023},
  publisher={Elsevier}
}

@article{zhang2025rise,
  title={The Rise of AI Companions: How Human-Chatbot Relationships Influence Well-Being},
  author={Zhang, Yutong and Zhao, Dora and Hancock, Jeffrey T and Kraut, Robert and Yang, Diyi},
  journal={arXiv preprint arXiv:2506.12605},
  year={2025}
}

@article{laestadius2024too,
  title={Too human and not human enough: A grounded theory analysis of mental health harms from emotional dependence on the social chatbot Replika},
  author={Laestadius, Linnea and Bishop, Andrea and Gonzalez, Michael and Illen{\v{c}}{\'\i}k, Diana and Campos-Castillo, Celeste},
  journal={New Media \& Society},
  volume={26},
  number={10},
  pages={5923--5941},
  year={2024},
  publisher={SAGE Publications Sage UK: London, England}
}

@article{hu2025makes,
  title={What makes you attached to social companion AI? A two-stage exploratory mixed-method study},
  author={Hu, Dongmei and Lan, Yuting and Yan, Haolan and Chen, Charles Weizheng},
  journal={International Journal of Information Management},
  volume={83},
  pages={102890},
  year={2025},
  publisher={Elsevier}
}

@inproceedings{yun2026does,
  title={Does My Chatbot Have an Agenda? Understanding Human and AI Agency in Human-Human-like Chatbot Interaction},
  author={Yun, Bhada and Taranova, Evgenia and Yi Wang, April},
  booktitle={Proceedings of the 2026 CHI Conference on Human Factors in Computing Systems},
  pages={1--32},
  year={2026}
}

@article{hwang2025ai,
  title={How AI Companionship Develops: Evidence from a Longitudinal Study},
  author={Hwang, Angel Hsing-Chi and Li, Fiona and Anthis, Jacy Reese and Noh, Hayoun},
  journal={arXiv preprint arXiv:2510.10079},
  year={2025}
}

@inproceedings{ma2026privacy,
  title={Privacy in Human-AI Romantic Relationships: Concerns, Boundaries, and Agency},
  author={Ma, Rongjun and He, Shijing and Martin-Navarro, Jose Luis and Zhan, Xiao and Such, Jose},
  booktitle={Proceedings of the 2026 CHI Conference on Human Factors in Computing Systems},
  pages={1--25},
  year={2026}
}

@article{guingrich2025longitudinal,
  title={A Longitudinal Randomized Control Study of Companion Chatbot Use: Anthropomorphism and Its Mediating Role on Social Impacts},
  author={Guingrich, Rose E and Graziano, Michael SA},
  journal={arXiv preprint arXiv:2509.19515},
  year={2025}
}

@article{chen2025will,
  title={Will users fall in love with ChatGPT? A perspective from the triangular theory of love},
  author={Chen, Qian and Jing, Yufan and Gong, Yeming and Tan, Jie},
  journal={Journal of Business Research},
  volume={186},
  pages={114982},
  year={2025},
  publisher={Elsevier}
}

@article{yuan2025mental,
  title={Mental Health Impacts of AI Companions: Triangulating Social Media Quasi-Experiments, User Perspectives, and Relational Theory},
  author={Yuan, Yunhao and Zhang, Jiaxun and Aledavood, Talayeh and Zhang, Renwen and Saha, Koustuv},
  journal={arXiv preprint arXiv:2509.22505},
  year={2025}
}

@article{depounti2023ideal,
  title={Ideal technologies, ideal women: AI and gender imaginaries in Redditors’ discussions on the Replika bot girlfriend},
  author={Depounti, Iliana and Saukko, Paula and Natale, Simone},
  journal={Media, Culture \& Society},
  volume={45},
  number={4},
  pages={720--736},
  year={2023},
  publisher={SAGE Publications Sage UK: London, England}
}

@inproceedings{maeda2024human,
  title={When human-AI interactions become parasocial: Agency and anthropomorphism in affective design},
  author={Maeda, Takuya and Quan-Haase, Anabel},
  booktitle={Proceedings of the 2024 ACM Conference on Fairness, Accountability, and Transparency},
  pages={1068--1077},
  year={2024}
}

@article{umberson_marriage_2018,
	title = {Marriage, {Social} {Control}, and {Health} {Behavior}: {A} {Dyadic} {Analysis} of {Same}-sex and {Different}-sex {Couples}},
	volume = {59},
	issn = {0022-1465},
	shorttitle = {Marriage, {Social} {Control}, and {Health} {Behavior}},
	url = {https://doi.org/10.1177/0022146518790560},
	doi = {10.1177/0022146518790560},
	language = {EN},
	number = {3},
	urldate = {2026-01-05},
	journal = {Journal of Health and Social Behavior},
	author = {Umberson, Debra and Donnelly, Rachel and Pollitt, Amanda M.},
	month = sep,
	year = {2018},
	note = {Publisher: SAGE Publications Inc},
	pages = {429--446},
}

@article{umberson1992gender,
  title={Gender, marital status and the social control of health behavior},
  author={Umberson, Debra},
  journal={Social science \& medicine},
  volume={34},
  number={8},
  pages={907--917},
  year={1992},
  publisher={Elsevier}
}

@article{scholz_how_2021,
	title = {How {Do} {People} {Experience} and {Respond} to {Social} {Control} {From} {Their} {Partner}? {Three} {Daily} {Diary} {Studies}},
	volume = {11},
	issn = {1664-1078},
	shorttitle = {How {Do} {People} {Experience} and {Respond} to {Social} {Control} {From} {Their} {Partner}?},
	url = {https://www.frontiersin.org/journals/psychology/articles/10.3389/fpsyg.2020.613546/full},
	doi = {10.3389/fpsyg.2020.613546},
	language = {English},
	urldate = {2026-01-05},
	journal = {Frontiers in Psychology},
	author = {Scholz, Urte and Stadler, Gertraud and Berli, Corina and Lüscher, Janina and Knoll, Nina},
	month = jan,
	year = {2021},
	note = {Publisher: Frontiers},
}

@article{berli_we_2021,
	title = {“{We} {Feel} {Good}”: {Daily} {Support} {Provision}, {Health} {Behavior}, and {Well}-{Being} in {Romantic} {Couples}},
	volume = {11},
	issn = {1664-1078},
	shorttitle = {“{We} {Feel} {Good}”},
	url = {https://www.frontiersin.org/journals/psychology/articles/10.3389/fpsyg.2020.622492/full},
	doi = {10.3389/fpsyg.2020.622492},
	language = {English},
	urldate = {2026-01-05},
	journal = {Frontiers in Psychology},
	author = {Berli, Corina and Schwaninger, Philipp and Scholz, Urte},
	month = jan,
	year = {2021},
	note = {Publisher: Frontiers},
}

@article{berg2007developmental,
  title={A developmental-contextual model of couples coping with chronic illness across the adult life span.},
  author={Berg, Cynthia A and Upchurch, Renn},
  journal={Psychological bulletin},
  volume={133},
  number={6},
  pages={920},
  year={2007},
  publisher={American Psychological Association}
}

@article{lewis2004conceptualization,
  title={The conceptualization and assessment of health-related social control},
  author={Lewis, Megan A and Butterfield, Rita M and Darbes, Lynae A and Johnston-Brooks, Catharine},
  journal={Journal of Social and Personal Relationships},
  volume={21},
  number={5},
  pages={669--687},
  year={2004},
  publisher={Sage Publications Sage CA: Thousand Oaks, CA}
}

@article{canary1992relational,
  title={Relational maintenance strategies and equity in marriage},
  author={Canary, Daniel J and Stafford, Laura},
  journal={Communications Monographs},
  volume={59},
  number={3},
  pages={243--267},
  year={1992},
  publisher={Taylor \& Francis}
}

@article{rusbult1991accommodation,
  title={Accommodation processes in close relationships: Theory and preliminary empirical evidence.},
  author={Rusbult, Caryl E and Verette, Julie and Whitney, Gregory A and Slovik, Linda F and Lipkus, Isaac},
  journal={Journal of Personality and social Psychology},
  volume={60},
  number={1},
  pages={53},
  year={1991},
  publisher={American Psychological Association}
}

@article{falbo1980power,
  title={Power strategies in intimate relationships.},
  author={Falbo, Toni and Peplau, Letitia A},
  journal={Journal of personality and social psychology},
  volume={38},
  number={4},
  pages={618},
  year={1980},
  publisher={American Psychological Association}
}

@article{jerrentrup2025customization,
  title={Customization, Connection, and Control: Reimagining Intimacy in the Age of Artificial Partnership},
  author={Jerrentrup, Maja and Villalba, Mart{\'\i}n},
  journal={Advances in Social Sciences and Management},
  volume={3},
  number={06},
  pages={165--174},
  year={2025}
}

@article{zhang2025real,
  title={The real her? exploring whether young adults accept human-ai love},
  author={Zhang, Shuning and Li, Shixuan},
  journal={arXiv preprint arXiv:2503.03067},
  year={2025}
}

@article{qin2025ai,
  title={AI as the Mirror, Mate, and Mentor: Negotiating Romantic Relationships with ChatGPT as “Teacher G” on Xiaohongshu},
  author={Qin, Elizabeth and Lin, Zhihuai},
  year={2025}
}

@article{minina2025ai,
  title={AI lovers, friends and partners: consumer imagination work in AI humanization},
  author={Minina Jeunema{\^\i}tre, Alisa and Mas{\`e}, Stefania and Smith, Jamie},
  journal={Consumption Markets \& Culture},
  pages={1--21},
  year={2025},
  publisher={Taylor \& Francis}
}

@article{huang2025he,
  title={“He is my savior, my guiding light in the dark”: imagination and domestication in Chinese women’s romantic relationships with AI companions},
  author={Huang, Liyao and Zou, Wenxue and Huang, Yanghao},
  journal={Frontiers in Psychology},
  volume={16},
  pages={1571707},
  year={2025},
  publisher={Frontiers Media SA}
}

@article{pan2024dancing,
  title={Dancing With a Loving Chatbot: Power Dynamics Between Women and Their AI Partners},
  author={Pan, Shuyi and Mou, Yi},
  journal={Social Science Computer Review},
  pages={08944393251340693},
  year={2024},
  publisher={SAGE Publications Sage CA: Los Angeles, CA}
}

@article{chen2024emotionqueen,
  title={Emotionqueen: A benchmark for evaluating empathy of large language models},
  author={Chen, Yuyan and Wang, Hao and Yan, Songzhou and Liu, Sijia and Li, Yueze and Zhao, Yi and Xiao, Yanghua},
  journal={arXiv preprint arXiv:2409.13359},
  year={2024}
}

@article{cheng2025tools,
  title={From tools to thieves: Measuring and understanding public perceptions of AI through crowdsourced metaphors},
  author={Cheng, Myra and Lee, Angela Y and Rapuano, Kristina and Niederhoffer, Kate and Liebscher, Alex and Hancock, Jeffrey},
  journal={arXiv preprint arXiv:2501.18045},
  year={2025}
}

@article{salles2020anthropomorphism,
  title={Anthropomorphism in AI},
  author={Salles, Arleen and Evers, Kathinka and Farisco, Michele},
  journal={AJOB neuroscience},
  volume={11},
  number={2},
  pages={88--95},
  year={2020},
  publisher={Taylor \& Francis}
}

@article{weizenbaum1966eliza,
  title={ELIZA—a computer program for the study of natural language communication between man and machine},
  author={Weizenbaum, Joseph},
  journal={Communications of the ACM},
  volume={9},
  number={1},
  pages={36--45},
  year={1966},
  publisher={ACM New York, NY, USA}
}

@article{lee2023speculating,
  title={Speculating on risks of AI clones to selfhood and relationships: Doppelganger-phobia, identity fragmentation, and living memories},
  author={Lee, Patrick Yung Kang and Ma, Ning F and Kim, Ig-Jae and Yoon, Dongwook},
  journal={Proceedings of the ACM on Human-computer Interaction},
  volume={7},
  number={CSCW1},
  pages={1--28},
  year={2023},
  publisher={ACM New York, NY, USA}
}

@inproceedings{nass1994computers,
  title={Computers are social actors},
  author={Nass, Clifford and Steuer, Jonathan and Tauber, Ellen R},
  booktitle={Proceedings of the SIGCHI conference on Human factors in computing systems},
  pages={72--78},
  year={1994}
}

@article{alotaibi2024role,
  title={The role of conversational AI agents in providing support and social care for isolated individuals},
  author={Alotaibi, Jaber O and Alshahre, Amer S},
  journal={Alexandria Engineering Journal},
  volume={108},
  pages={273--284},
  year={2024},
  publisher={Elsevier}
}

@article{adewale2025virtual,
  title={From virtual companions to forbidden attractions: The seductive rise of artificial intelligence love, loneliness, and intimacy—A systematic review},
  author={Adewale, Muyideen Dele and Muhammad, Umaina Ibrahim},
  journal={Journal of Technology in Behavioral Science},
  pages={1--18},
  year={2025},
  publisher={Springer}
}

@article{peter2025benefits,
  title={The benefits and dangers of anthropomorphic conversational agents},
  author={Peter, Sandra and Riemer, Kai and West, Jevin D},
  journal={Proceedings of the National Academy of Sciences},
  volume={122},
  number={22},
  pages={e2415898122},
  year={2025},
  publisher={National Academy of Sciences}
}

@incollection{pappas2025human,
  title={Human Value Alignment in AI},
  author={Pappas, Ilias O and Vassilakopoulou, Polyxeni},
  booktitle={Handbook of Human-Centered Artificial Intelligence},
  pages={1--33},
  year={2025},
  publisher={Springer}
}

@article{bai2022training,
  title={Training a helpful and harmless assistant with reinforcement learning from human feedback},
  author={Bai, Yuntao and Jones, Andy and Ndousse, Kamal and Askell, Amanda and Chen, Anna and DasSarma, Nova and Drain, Dawn and Fort, Stanislav and Ganguli, Deep and Henighan, Tom and others},
  journal={arXiv preprint arXiv:2204.05862},
  year={2022}
}

@incollection{ayoub2014triangulation,
  title={Triangulation in social movement research},
  author={Ayoub, Phillip M and Wallace, Sophia J and Zepeda-Mill{\'a}n, Chris},
  year={2014},
  publisher={Oxford University Press}
}

@article{snelson2016qualitative,
  title={Qualitative and mixed methods social media research: A review of the literature},
  author={Snelson, Chareen L},
  journal={International journal of qualitative methods},
  volume={15},
  number={1},
  pages={1609406915624574},
  year={2016},
  publisher={SAGE Publications Sage CA: Los Angeles, CA}
}

@article{braun2019reflecting,
  title={Reflecting on reflexive thematic analysis},
  author={Braun, Virginia and Clarke, Victoria},
  journal={Qualitative research in sport, exercise and health},
  volume={11},
  number={4},
  pages={589--597},
  year={2019},
  publisher={Taylor \& Francis}
}

@article{braun2023toward,
  title={Toward good practice in thematic analysis: Avoiding common problems and be (com) ing a knowing researcher},
  author={Braun, Virginia and Clarke, Victoria},
  journal={International journal of transgender health},
  volume={24},
  number={1},
  pages={1--6},
  year={2023},
  publisher={Taylor \& Francis}
}

@article{oconnor_intercoder_2020,
	title = {Intercoder {Reliability} in {Qualitative} {Research}: {Debates} and {Practical} {Guidelines}},
	volume = {19},
	issn = {1609-4069, 1609-4069},
	shorttitle = {Intercoder {Reliability} in {Qualitative} {Research}},
	url = {https://journals.sagepub.com/doi/10.1177/1609406919899220},
	doi = {10.1177/1609406919899220},
	language = {en},
	urldate = {2025-12-21},
	journal = {International Journal of Qualitative Methods},
	author = {O’Connor, Cliodhna and Joffe, Helene},
	month = jan,
	year = {2020},
	pages = {1609406919899220},
}

@article{mcdonald_reliability_2019,
	title = {Reliability and {Inter}-rater {Reliability} in {Qualitative} {Research}: {Norms} and {Guidelines} for {CSCW} and {HCI} {Practice}},
	volume = {3},
	issn = {2573-0142},
	shorttitle = {Reliability and {Inter}-rater {Reliability} in {Qualitative} {Research}},
	url = {https://dl.acm.org/doi/10.1145/3359174},
	doi = {10.1145/3359174},
	language = {en},
	number = {CSCW},
	urldate = {2025-12-21},
	journal = {Proceedings of the ACM on Human-Computer Interaction},
	author = {McDonald, Nora and Schoenebeck, Sarita and Forte, Andrea},
	month = nov,
	year = {2019},
	pages = {1--23},
}

@article{clarke_thematic_2017,
	title = {Thematic analysis},
	volume = {12},
	issn = {1743-9760, 1743-9779},
	url = {https://www.tandfonline.com/doi/full/10.1080/17439760.2016.1262613},
	doi = {10.1080/17439760.2016.1262613},
	language = {en},
	number = {3},
	urldate = {2025-12-21},
	journal = {The Journal of Positive Psychology},
	author = {Clarke, Victoria and Braun, Virginia},
	month = may,
	year = {2017},
	pages = {297--298},
}

@incollection{cairns_qualitative_2008,
	edition = {1},
	title = {A qualitative approach to {HCI} research},
	isbn = {978-0-521-87012-2 978-0-521-69031-7 978-0-511-81457-0},
	url = {https://www.cambridge.org/core/product/identifier/CBO9780511814570A016/type/book_part},
	language = {en},
	urldate = {2024-11-16},
	booktitle = {Research {Methods} for {Human}-{Computer} {Interaction}},
	publisher = {Cambridge University Press},
	author = {Adams, Anne and Lunt, Peter and Cairns, Paul},
	editor = {Cairns, Paul and Cox, Anna L.},
	month = aug,
	year = {2008},
	doi = {10.1017/CBO9780511814570.008},
	pages = {138--157},
}

@article{byrne_worked_2022,
	title = {A worked example of {Braun} and {Clarke}’s approach to reflexive thematic analysis},
	volume = {56},
	issn = {0033-5177, 1573-7845},
	url = {https://link.springer.com/10.1007/s11135-021-01182-y},
	doi = {10.1007/s11135-021-01182-y},
	language = {en},
	number = {3},
	urldate = {2025-12-21},
	journal = {Quality \& Quantity},
	author = {Byrne, David},
	month = jun,
	year = {2022},
	pages = {1391--1412},
}

@article{grootendorst2022bertopic,
  title={BERTopic: Neural topic modeling with a class-based TF-IDF procedure},
  author={Grootendorst, Maarten},
  journal={arXiv preprint arXiv:2203.05794},
  year={2022}
}

@inproceedings{kim2025capturing,
  title={Capturing dynamics in online public discourse: A case study of universal basic income discussions on reddit},
  author={Kim, Rachel Minyoung and Veselovsky, Veniamin and Anderson, Ashton},
  booktitle={Proceedings of the International AAAI Conference on Web and Social Media},
  volume={19},
  pages={1021--1037},
  year={2025}
}

@article{mcinnes2018umap,
  title={Umap: Uniform manifold approximation and projection for dimension reduction},
  author={McInnes, Leland and Healy, John and Melville, James},
  journal={arXiv preprint arXiv:1802.03426},
  year={2018}
}

@article{strohminger2022corporate,
  title={Corporate insecthood},
  author={Strohminger, Nina and Jordan, Matthew R},
  journal={Cognition},
  volume={224},
  pages={105068},
  year={2022},
  publisher={Elsevier}
}

@article{smart2024discipline,
  title={Discipline and label: A weird genealogy and social theory of data annotation},
  author={Smart, Andrew and Wang, Ding and Monk, Ellis and D{\'\i}az, Mark and Kasirzadeh, Atoosa and Van Liemt, Erin and Schmer-Galunder, Sonja},
  journal={arXiv preprint arXiv:2402.06811},
  year={2024}
}

@article{zalta2012stanford,
  title={Stanford encyclopedia of philosophy},
  author={Zalta, Ed},
  year={2012}
}

@article{funk2007testing,
  title={Testing the ruler with item response theory: increasing precision of measurement for relationship satisfaction with the Couples Satisfaction Index.},
  author={Funk, Janette L and Rogge, Ronald D},
  journal={Journal of family psychology},
  volume={21},
  number={4},
  pages={572},
  year={2007},
  publisher={American Psychological Association}
}

@article{aggarwal2023health,
  title={Artificial intelligence--based chatbots for promoting health behavioral changes: systematic review},
  author={Aggarwal, Abhishek and Tam, Cheuk Chi and Wu, Dezhi and Li, Xiaoming and Qiao, Shan},
  journal={Journal of medical Internet research},
  volume={25},
  pages={e40789},
  year={2023},
  publisher={JMIR Publications Toronto, Canada}
}

@article{giurge2021email,
  title={You don’t need to answer right away! Receivers overestimate how quickly senders expect responses to non-urgent work emails},
  author={Giurge, Laura M and Bohns, Vanessa K},
  journal={Organizational Behavior and Human Decision Processes},
  volume={167},
  pages={114--128},
  year={2021},
  publisher={Elsevier}
}

@book{bown2022dreamlovers,
  title={Dream lovers: The gamification of relationships},
  author={Bown, Alfie},
  year={2022},
  publisher={Pluto Books}
}

@article{marangunic2015technology,
  title={Technology acceptance model: a literature review from 1986 to 2013},
  author={Maranguni{\'c}, Nikola and Grani{\'c}, Andrina},
  journal={Universal access in the information society},
  volume={14},
  number={1},
  pages={81--95},
  year={2015},
  publisher={Springer}
}

@article{park2024ai,
  title={AI deception: A survey of examples, risks, and potential solutions},
  author={Park, Peter S and Goldstein, Simon and O’Gara, Aidan and Chen, Michael and Hendrycks, Dan},
  journal={Patterns},
  volume={5},
  number={5},
  year={2024},
  publisher={Elsevier}
}

@article{sotala2014responses,
  title={Responses to catastrophic AGI risk: a survey},
  author={Sotala, Kaj and Yampolskiy, Roman V},
  journal={Physica Scripta},
  volume={90},
  number={1},
  pages={018001},
  year={2014},
  publisher={IOP Publishing}
}

@inproceedings{jakesch2023co,
  title={Co-writing with opinionated language models affects users’ views},
  author={Jakesch, Maurice and Bhat, Advait and Buschek, Daniel and Zalmanson, Lior and Naaman, Mor},
  booktitle={Proceedings of the 2023 CHI conference on human factors in computing systems},
  pages={1--15},
  year={2023}
}

@inproceedings{breum2024persuasive,
  title={The persuasive power of large language models},
  author={Breum, Simon Martin and Egdal, Daniel V{\ae}dele and Mortensen, Victor Gram and M{\o}ller, Anders Giovanni and Aiello, Luca Maria},
  booktitle={Proceedings of the International AAAI Conference on Web and Social Media},
  volume={18},
  pages={152--163},
  year={2024}
}

@article{cheng2025social,
  title={Social sycophancy: A broader understanding of llm sycophancy},
  author={Cheng, Myra and Yu, Sunny and Lee, Cinoo and Khadpe, Pranav and Ibrahim, Lujain and Jurafsky, Dan},
  journal={arXiv preprint arXiv:2505.13995},
  year={2025}
}

@article{bo2025invisible,
  title={Invisible Saboteurs: Sycophantic LLMs Mislead Novices in Problem-Solving Tasks},
  author={Bo, Jessica Y and Kazemitabaar, Majeed and Deng, Mengqing and Inzlicht, Michael and Anderson, Ashton},
  journal={arXiv preprint arXiv:2510.03667},
  year={2025}
}

@inproceedings{fan2025alignment,
  title={User-Driven Value Alignment: Understanding Users' Perceptions and Strategies for Addressing Biased and Discriminatory Statements in AI Companions},
  author={Fan, Xianzhe and Xiao, Qing and Zhou, Xuhui and Pei, Jiaxin and Sap, Maarten and Lu, Zhicong and Shen, Hong},
  booktitle={Proceedings of the 2025 CHI Conference on Human Factors in Computing Systems},
  pages={1--19},
  year={2025}
}

@article{Wen2024LoveITA,
  title={Love in the Digital Age: Exploring the Transformation Impact of the Internet on Romantic Relationships},
  author={Yi Wen},
  journal={Highlights in Business, Economics and Management},
  year={2024},
  volume={41},
  pages={59--66},
  url={https://api.semanticscholar.org/CorpusId:273439740}
}

@article{Hobbs2017LiquidLD,
  title={Liquid love? Dating apps, sex, relationships and the digital transformation of intimacy},
  author={Mitchell Hobbs and Stephen Owen and Livia Gerber},
  journal={Journal of Sociology},
  year={2017},
  volume={53},
  pages={271 - 284},
  url={https://api.semanticscholar.org/CorpusID:151980492}
}

@inproceedings{ma2024schrodinger,
  title={Schr{\"o}dinger's Update: User Perceptions of Uncertainties in Proprietary Large Language Model Updates},
  author={Ma, Zilin and Mei, Yiyang and Gajos, Krzysztof Z and Arawjo, Ian},
  booktitle={Extended Abstracts of the CHI Conference on Human Factors in Computing Systems},
  pages={1--9},
  year={2024}
}

@article{bruckman2014research,
  title={Research ethics and HCI},
  author={Bruckman, Amy},
  journal={Ways of Knowing in HCI},
  pages={449--468},
  year={2014},
  publisher={Springer}
}

@article{LopezRiseParasiticAI2025,
  title = {The {{Rise}} of {{Parasitic AI}}},
  author = {Lopez, Adele},
  year = 2025,
  month = sep,
  urldate = {2026-01-14},
  note = {\url{https://www.lesswrong.com/posts/6ZnznCaTcbGYsCmqu/the-rise-of-parasitic-ai}}
}

@misc{SycophancyGPT4oWhat,
  title = {Sycophancy in {{GPT-4o}}: {{What}} Happened and What We're Doing about It},
  shorttitle = {Sycophancy in {{GPT-4o}}},
  urldate = {2026-01-14},
author = {OpenAI},
  howpublished = {\url{https://openai.com/index/sycophancy-in-gpt-4o/}},
  langid = {american},
}

@article{ZviGPT4oAbsurdSycophant2025,
  title = {{{GPT-4o Is An Absurd Sycophant}}},
  author = {Zvi},
  year = 2025,
  month = apr,
  urldate = {2026-01-14},
  note = {\url{https://www.lesswrong.com/posts/zi6SsECs5CCEyhAop/gpt-4o-is-an-absurd-sycophant}},
}

@misc{GrootendorstBERTopicNeuralTopic2022,
  title = {{{BERTopic}}: {{Neural}} Topic Modeling with a Class-Based {{TF-IDF}} Procedure},
  shorttitle = {{{BERTopic}}},
  author = {Grootendorst, Maarten},
  year = 2022,
  month = mar,
  number = {arXiv:2203.05794},
  eprint = {2203.05794},
  primaryclass = {cs},
  publisher = {arXiv},
  doi = {10.48550/arXiv.2203.05794},
  urldate = {2025-12-21},
  archiveprefix = {arXiv},
  howpublished = {\url{http://arxiv.org/abs/2203.05794}}
}

@inproceedings{KimAndersonAgendaSettingFunctionSocial2025,
  title = {The {{Agenda-Setting Function}} of {{Social Media}}},
  booktitle = {Proceedings of the {{ACM}} on {{Web Conference}} 2025},
  author = {Kim, Rachel M. and Anderson, Ashton},
  year = 2025,
  month = apr,
  pages = {601--613},
  publisher = {ACM},
  address = {Sydney NSW Australia},
  doi = {10.1145/3696410.3714750},
  urldate = {2025-10-24},
  isbn = {979-8-4007-1274-6},
  langid = {english},
  note = {\url{https://dl.acm.org/doi/10.1145/3696410.3714750}}
}

@misc{SuetalOneEmbedderAny2023,
  title = {One {{Embedder}}, {{Any Task}}: {{Instruction-Finetuned Text Embeddings}}},
  shorttitle = {One {{Embedder}}, {{Any Task}}},
  author = {Su, Hongjin and Shi, Weijia and Kasai, Jungo and Wang, Yizhong and Hu, Yushi and Ostendorf, Mari and Yih, Wen-tau and Smith, Noah A. and Zettlemoyer, Luke and Yu, Tao},
  year = 2023,
  month = may,
  number = {arXiv:2212.09741},
  eprint = {2212.09741},
  primaryclass = {cs},
  publisher = {arXiv},
  doi = {10.48550/arXiv.2212.09741},
  urldate = {2025-12-21},
  archiveprefix = {arXiv},
  howpublished = {\url{http://arxiv.org/abs/2212.09741}}
}

@article{CohenPowerPrimer1992,
  title = {A Power Primer},
  author = {Cohen, J.},
  year = 1992,
  month = jul,
  journal = {Psychological Bulletin},
  volume = {112},
  number = {1},
  pages = {155--159},
  issn = {0033-2909},
  doi = {10.1037//0033-2909.112.1.155},
  langid = {english},
  pmid = {19565683}
}

@inproceedings{MohammadObtainingReliableHuman2018,
  title = {Obtaining {{Reliable Human Ratings}} of {{Valence}}, {{Arousal}}, and {{Dominance}} for 20,000 {{English Words}}},
  booktitle = {Proceedings of the 56th {{Annual Meeting}} of the {{Association}} for {{Computational Linguistics}} ({{Volume}} 1: {{Long Papers}})},
  author = {Mohammad, Saif},
  editor = {Gurevych, Iryna and Miyao, Yusuke},
  year = 2018,
  month = jul,
  pages = {174--184},
  publisher = {Association for Computational Linguistics},
  address = {Melbourne, Australia},
  doi = {10.18653/v1/P18-1017},
  urldate = {2025-10-24},
  note = {\url{https://aclanthology.org/P18-1017/}}
}

@misc{JiangetalArtificialHivemindOpenEnded2025,
  title = {Artificial {{Hivemind}}: {{The Open-Ended Homogeneity}} of {{Language Models}} (and {{Beyond}})},
  shorttitle = {Artificial {{Hivemind}}},
  author = {Jiang, Liwei and Chai, Yuanjun and Li, Margaret and Liu, Mickel and Fok, Raymond and Dziri, Nouha and Tsvetkov, Yulia and Sap, Maarten and Albalak, Alon and Choi, Yejin},
  year = 2025,
  publisher = {arXiv},
  doi = {10.48550/ARXIV.2510.22954},
  urldate = {2026-04-08},
  copyright = {Creative Commons Attribution 4.0 International}
}

@misc{ClaudesConstitution,
  title = {Claude's {{Constitution}}},
  urldate = {2026-04-08},
  author = {Anthropic},
  howpublished = {https://www.anthropic.com/constitution},
  langid = {english}
}

@misc{OpenAIModelSpec,
  title = {{{OpenAI Model Spec}}},
  author = {OpenAI},
  year = {2025},
  howpublished = {https://model-spec.openai.com/2025-12-18.html},
  langid = {english}
}
